\documentclass[structabstract]{aa}  
\usepackage[fleqn]{amsmath}
\usepackage{amsbsy}
\usepackage{graphicx}
\usepackage{txfonts}
\usepackage{natbib}
\usepackage{caption}
\bibpunct{(}{)}{;}{a}{}{,} 

\newcommand{\bsigma}{\sigma \kern -0.7em \sigma}

\begin{document}

\authorrunning{P. Noorishad et al.}
\title{Redundancy Calibration of Phased Array Stations}

\subtitle{}

  \author{P. Noorishad
          \inst{1,}\inst{2},
          S. J. Wijnholds
	  \inst{2},
	  A. van Ardenne
	  \inst{2,}\inst{3}
	  \and
	  J. M. van der Hulst 
	  \inst{1}
          }

\institute{Kapteyn Astronomical Institute- University of Groningen,
               Landleven 12, 9747 AV, Groningen, The Netherlands\\
         \and
              ASTRON- Oude Hoogeveensedijk 4, 7991 PD Dwingeloo, The Netherlands\\
	 \and
	      Chalmers University of Technology- SE-412 96 Gothenburg, Sweden}

\date{Accepted: 1 May 2012}

\abstract
{}
{Our aim is to assess the benefits and limitations of using the redundant visibility information in regular phased array systems for improving the calibration.}
{Regular arrays offer the possibility to use redundant visibility information to constrain the calibration of the array independent of a sky model and a beam models of the station elements. It requires a regular 
arrangement in the configuration of array elements and identical beam patterns. }
{We revised a calibration method for phased array stations using the redundant visibility information in the system  and applied it successfully to a LOFAR station. The performance and limitations of the method were demonstrated by comparing its use on real and simulated data.
The main limitation is the mutual coupling between the station elements, which leads to non-identical beams and stronger baseline dependent noise. Comparing the variance of the estimated 
complex gains with the Cramer-Rao Bound (CRB) indicates that redundancy is a stable and optimum method for calibrating the complex gains of the system.}
{ Our study shows that the use of  the redundant visibility does improve the quality of the calibration in phased array systems. In addition it provides a powerful tool for system diagnostics.  Our results demonstrate that designing redundancy in both the station layout and the array configuration of future aperture arrays is strongly recommended. In particular in the case of the Square Kilometre Array with its dynamic range requirement which surpasses any existing array by an order of magnitude. 
}

\maketitle

\section{Introduction}
\label{Intro.}

\begin{figure}
\centering
\includegraphics [width=6cm] {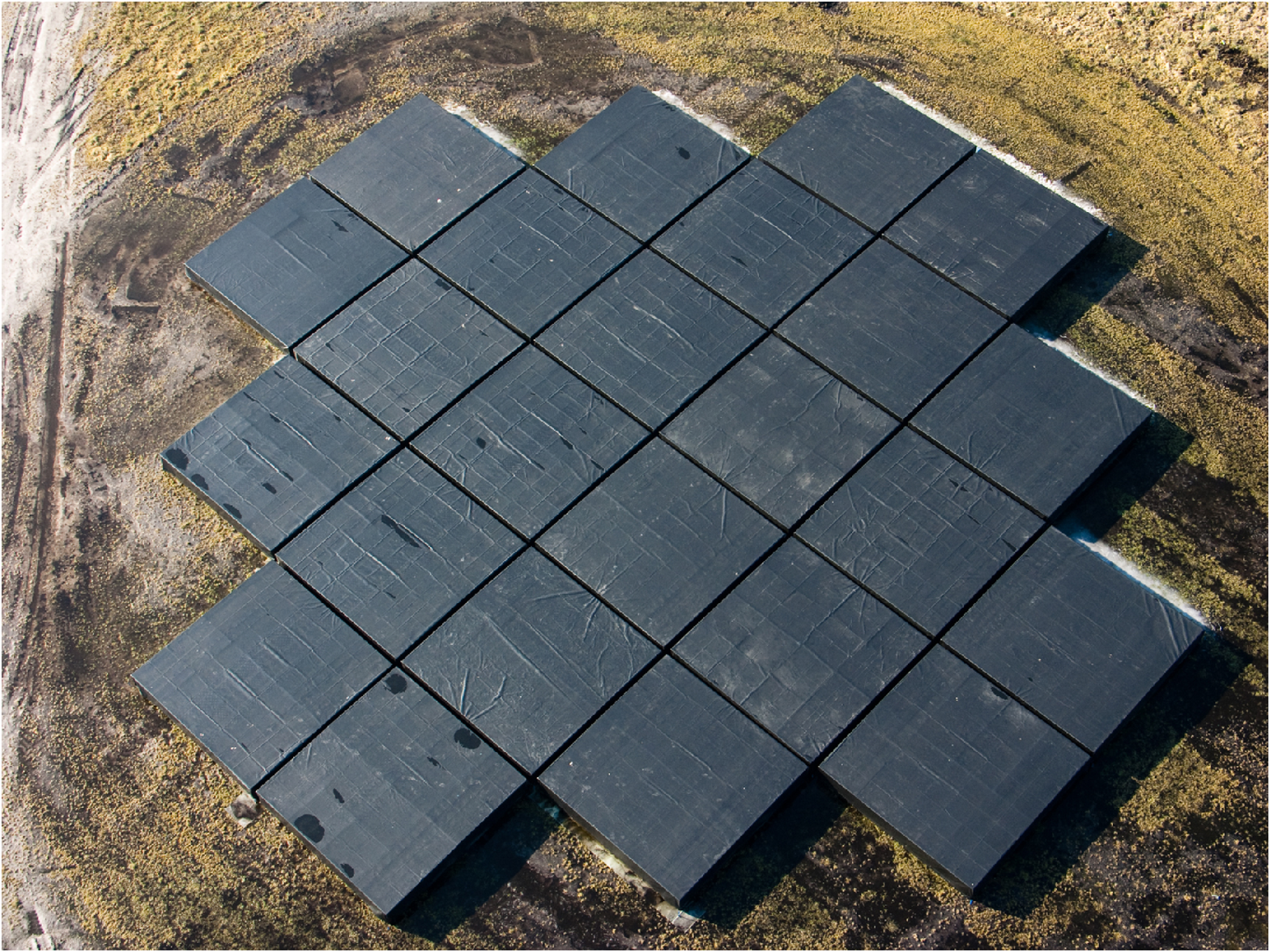}
\caption{24-tile HBA station. Each tile is one station element. This is the station configuration for most of LOFAR's HBA stations including the LOFAR core or CS302. It is clear that a station like this is highly
redundant.}
\label{Fig2}
\end{figure}

An important conceptual difference between the next generation of radio telescopes and conventional ones is their hierarchical system architecture. An excellent example is the Low Frequency ARray (LOFAR),
 \citet{2009IEEEP..97.1431D}. In LOFAR, a station consists of phased arrays, either sparse or dense. In the sparse phased array stations, the station elements are dipoles which are digitally 
beamformed to synthesize a station as a dish such as the LBAs (Low Band Antennas) in a LOFAR station. Phased arrays operating above $\sim 100$ MHz are often implemented as compound elements or tiles such as the HBAs
 (High Band Antennas) at the LOFAR stations or in EMBRACE (Electronic Multi-Beam Radio Astronomy ConcEpt), \citet{2004ExA....17...65A}; \citet{Kant2011}. EMBRACE is an example of a very dense phased array
 station. In these stations, the station elements are phased array tiles (e.g. Fig. \ref{Fig2}). A tile is a regular arrangement of many dipoles whose signals are added in phase to form an instantaneous beam (the beamforming). The tile output signals are digitally phased up to 
synthesize a station as a dish. At the next level in the beamforming hierarchy, the beam formed output of each station is transported to the central correlator to synthesize the whole telescope. 
Calibration has to be performed at different levels of this hierarchy to provide a final, high dynamic range image of part of sky as explained by \citet{2010ISPM...27...30W}. 

In this paper, we are concentrate on the calibration at station level. Its purpose is to ensure the station beam stability over time and frequency. A robust calibration as part of the beamforming process 
should guarantee a stable beam pattern of the station for data going to the central correlator. This is crucial for the dynamic range of the final images made using data from the entire array.

\begin{figure}
\centering
\includegraphics [width=9cm] {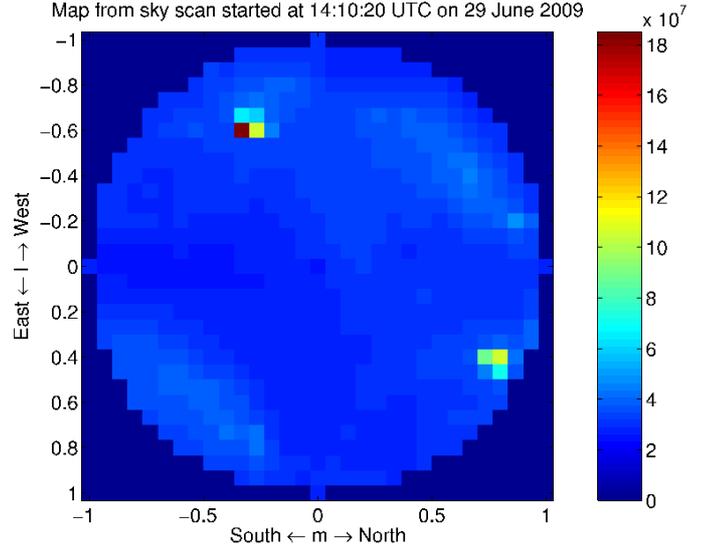}
\caption{The sky imaged by HBA tiles at 14:10:20 UTC on 29 June 2009. The Galactic plane appears in the north-west, the sun appears in the south-west. One can also see their corresponding grating response in
the image. The image is presented in $(l,m)$-coordinates. $l= cos(el) sin(az)$ and $m = cos(el) cos(az)$, where $el$ and $az$ denote elevation and azimuth respectively.}
\label{Fig1}
\end{figure}

In a phased array station, the output of all station elements can be correlated. These correlations are called station visibilities which are used for engineering purposes e.g. station calibration, RFI detection/mitigation. These correlations include many short baselines on which extended structures such as the galactic plane are captured. The most commonly used calibration methods for phased array stations are model based, such as a multi-source calibration method introduced by \citet{2009ITSP...57.3512W}. A model based calibration method requires the presence of one or more relatively unresolved point sources such as CasA and a model of the extended structures (see Fig.\ref{Fig1}). Modeling an extended structure is computationally difficult and expensive. \citet{2010arXiv1003.2497W} phenomenologically model it as correlated noise and estimate the parameters of interest for calibration using a WALS (Weighted Alternative Least Squares) approach. However, the model-based methods are in general iterative methods, which usually converge after several iterations.

A regular arrangement of station elements has the advantage that it provides redundant baselines, i.e. baselines with the same physical length and orientation. Using redundant baseline information for calibration was introduced by \citet{1982Natur.299..597N}. Its linearity, independence of a sky model, low computational cost and proven efficiency for precision calibration of WSRT (Westerbork Synthesis Radio Telescope) observations motivated us to apply the redundancy calibration to phased array stations. The redundancy calibration algorithm uses the data of all redundant baselines to obtain a convergent calibration solution in a single step.

However, redundancy calibration in phased array systems requires additional considerations. This is essentially because of their different design concepts e.g. the closely located elements of a phased array station experience mutual coupling between elements which leads to non-identical beams of the station elements and to correlated receiver noise. 

In this paper, we refine the standard data model presented in the phased array signal processing literature to account for baseline dependent corruptions in terms of the coupling effects. Using this refined data model, we briefly introduce the two calibration methods i.e. model based and redundancy. This helps us to achieve a better understanding of the potential and limitations of both calibration methods. We will also revise the redundancy method formalism presented by (\citet{Wieringa1991}; \citet{2010MNRAS.408.1029L}) to capture the nature of baseline dependent errors which affect the calibration accuracy. Some implementation issues will be raised and investigated using observed and simulated data of LOFAR HBA stations. We will evaluate the redundancy calibration performance by comparing the variance of its results with the CRB (Cramer-Rao Bound) and the plots of residuals for the corrected data after redundancy calibration. We will also discuss limiting factors for its applicability.

Although we have used HBA data to demonstrate the applicability and efficiency of the redundancy method, the analysis in this paper will be relevant to any phased array which is to be calibrated using redundant visibility information.

Notation: we denote vectors in bold lowercase letters and matrices in bold uppercase letters. The matrix transpose and Hermitian transpose are denoted by $(.)^T$ and $(.)^H$ respectively. 
Operator $\mathrm{diag}(.)$ creates a diagonal matrix of a given vector. Operator $\mathrm{vec}(.)$ creates a vector. Operator $\odot$ denotes element-wise matrix multiplication. 
Operator $\angle(.)$ returns the angle or phase of a complex number. Operator $\kappa(.)$ returns the condition number of a matrix.

\section{Methods}
\label{Meth.}

\subsection{Data model for phased arrays}
\label{DaMod.}

The standard data model for phased array stations presented in the literature assumes that in the absence of RFI and any coupling effects, a phased array of $p$ elements has a signal vector,
 $\mathbf{x}(t)=[x_1(t), x_2(t), ... ,x_p(t)]^T$ which can be expressed as:

\begin{equation}
\mathbf{x}(t)=\mathbf{\Gamma} \mathbf{\Phi} \left(\sum_{k=1}^{q} \mathbf{a}_{k} s_{k}(t)\right) +\mathbf{n}(t)=\mathbf{\Gamma} \mathbf{\Phi} \mathbf{A} \mathbf{s}(t)+\mathbf{n}(t)
\label{eq:DM1}
\end{equation}

\noindent where $\mathbf{s}(t)$ is a $q \times 1$ vector containing $q$ mutually independent i.i.d.\footnote{temporally independent and identically distributed.} Gaussian signals impinging on the array with the
 covariance of $\mathbf{\Sigma}_{\mathrm{s}}=\mathrm{diag}(\boldsymbol{\sigma}_s)$ (size $q \times q$), where $\boldsymbol{\sigma}_s$ is a vecor of the source fluxes. They are also assumed to be narrow band, so we
 can define the $q$ spatial signature vectors $\mathbf{a}_{k}$ which include the phase delays due to the geometry and the directional response of the receiving elements (assumed to be identical).
 The vectors $\mathbf{a}_{k}$ are called the array response vectors which are usually normalized. The receiver noise signals $n_{i}(t)$ are assumed to be mutually independent i.i.d. Gaussian signals in a
 $p \times 1$ vector $\mathbf{n}(t)$ and uncorrelated. Thus, the noise covariance matrix, $\mathbf{\Sigma}_{\mathrm{n}}=\mathrm{diag}({{\boldsymbol{\sigma}}}_n)$ (size $p \times p$). The amplitudes and the phases
 of direction independent complex gains ($g_i$'s) which have to be calibrated are $\boldsymbol{\gamma}=[\gamma_1,\gamma_2,...,\gamma_p]^T$ and $\boldsymbol{\phi}=[e^{j\phi_1}, e^{j\phi_2}, ..., e^{j\phi_p}]^T$. 
Correspondingly $\mathbf {\Gamma}=\mathrm{diag}(\boldsymbol{\gamma})$ and $\mathbf{\Phi}=\mathrm{diag}(\boldsymbol{\phi})$. $\mathbf{A}=[\mathbf{a}_1,\mathbf{a}_2,...,\mathbf{a}_q]$ (size $p \times q$) is a stack
 of the array response vectors. Before computing the coherency, $\mathbf{x}(t)$ is sampled with period $T$. The $n$th sample of the array signal vector $\mathbf{x}[n]$ is given by:

\begin{equation}
\mathbf{x}[n] = \sum_{-\infty}^{\infty} \mathbf{x}(t) \delta(t-nT) = \mathbf{x}(nT)
\end{equation}

$N$ samples can be stacked in a matrix $\mathbf{X} = [\mathbf{x}[1], \mathbf{x}[2], ...,\mathbf{x}[N]]$ (size $p \times N$) which denotes the short term integrtion data set or snapshot. 
The array covariance matrix or the visibility matrix describing the correlation between all sampled volatges can be estimated by $\hat{\mathbf{R}} = \mathbf{X}\mathbf{X}^T/N$ whose expected value becomes:

\begin{equation}
\mathbf{R}=\mathbf{\Gamma}\mathbf{\Phi}\mathbf{A}\mathbf{\Sigma}_{\mathrm{s}}\mathbf{A}^H\mathbf{\Phi}^H\mathbf{\Gamma}^H+\mathbf{\Sigma}_{\mathrm{n}}
\label{eq:DM2}
\end{equation}

\noindent or,

\begin{equation}
\mathbf{R}=\mathbf{G}\mathbf{A}\mathbf{\Sigma}_{\mathrm{s}}\mathbf{A}^H \mathbf{G}^H+\mathbf{\Sigma}_{\mathrm{n}}
\label{eq:DM3}
\end{equation}

\noindent Correlator errors can be represented as an additive term in the covariance matrix i.e. as a non-diagonal matrix. However in Eq. \ref{eq:DM2} and throughout this paper, they are disregarded assuming that the correlator is designed perfectly.

Note that $\gamma_i$ could accommodate the overall amplitude gain of both the receiver system and the atmospheric disturbances and $\phi_i$ the corresponding phase shift. In the case of station calibration,
we do not calibrate for direction dependent effects because we use snapshot data. In model-based methods, these are absorbed in the known sky. In the case of redundancy calibration, these are absorbed in its fundamental assumption. We will elaborate on this assumption  in Sect. \ref{RedunMeth}. 

In phased arrays such as the HBA stations of LOFAR, or EMBRACE, the antenna elements are closely packed. This may cause mutual coupling between them, i.e. not all the power received by the elements is 
absorbed but some of the power is reradiated to the other elements. The reradiated power induces new currents in the other elements. Consequently, the radiation pattern of the elements changes. This leads to
non-identicalness of the element beam patterns or different array response vectors despite their physical identicalness (as it is considered in $\mathbf{A}$). This effect can be  modeled as a direction dependent
gain, $\mathbf{G}_0$ (size $p \times q$) being element-wise multiplied by the array response vectors:

\begin{equation}
\mathbf{R}=\mathbf{G}(\mathbf{G}_0 \odot \mathbf{A})\mathbf{\Sigma}_{\mathrm{s}}(\mathbf{G}_0 \odot \mathbf{A})^H\mathbf{M}^H\mathbf{G}^H+\mathbf{\Sigma}_{\mathrm{n}}
\label{eq:DM4_2}
\end{equation}

As it was studied by \citet{Svantesson1988}, mutual coupling can directly be represented by $\mathbf{M}$ (size $p \times p$) in the data model:

\begin{equation}
\mathbf{R}=\mathbf{G}\mathbf{M}\mathbf{A}\mathbf{\Sigma}_{\mathrm{s}}\mathbf{A}^H\mathbf{M}^H\mathbf{G}^H+\mathbf{\Sigma}_{\mathrm{n}}
\label{eq:DM4}
\end{equation}

\noindent One may notice that a direct association of $\mathbf{M}$ and $\mathbf{G}_0$ can not be expressed analytically. It requires a numerical evaluation which we present in Sect.\ref{RedunChek.}. For simplicity in order to continue our 
argument, we use the data model in Eq. \ref{eq:DM4}. 

Mutual coupling may not only act on the signal, but also on the system noise. The LNA's (Low Noise Amplifiers) connected to the antennas in such an array, generate EM noise waves towards their outputs,
but also send EM noise waves back into the antenna array. These waves are coupled into other receiver channels, giving rise to a correlated noise contribution. This effect is known as noise coupling which contributes
as $\mathbf{R}_{rec}$ in a general and non-diagonal noise correlation matrix represented as: 

\begin{equation}
\mathbf{\Sigma}'_{\mathrm{n}} = \mathbf{R}_{sp} + \mathbf{R}_{sky} + \mathbf{R}_{rec} 
\label{eq:DM5}
\end{equation}

\noindent where $\mathbf{R}_{sp}$ is the spillover noise correlation matrix which can usually be ignored as compared with $\mathbf{R}_{sky}$ which is the sky noise contribution. Crosstalk in the back-end adds another baseline
specific correlated noise term which we disregard here. Then, the general data model for the visibility matrix becomes:

\begin{equation}
\mathbf{R}=\mathbf{G}\mathbf{M}\mathbf{A}\mathbf{\Sigma}_{\mathrm{s}}\mathbf{A}^H\mathbf{M}^H\mathbf{G}^H+\mathbf{\Sigma}'_{\mathrm{n}}
\label{eq:DM6}
\end{equation}
 
Note that $\mathbf{\Sigma}'_{\mathrm{n}}$ is not a diagonal matrix, unlike $\mathbf{\Sigma}_{\mathrm{n}}$. It can be shown that noise from each station element has still a normal distribution in view of large number of samples and according to the central limit theorem. Therefore, the off-diagonal elements of $\mathbf{\Sigma}'_{\mathrm{n}}$ which will appear in Eq. \ref{eq:Redun1} have a Wishart distribution.

Although the noise model in Eq. \ref{eq:DM5} has been presented only in  literature concerning PAFs (Phased Array Feeds), for example by \citet{2008ISTSP...2..635J} and \citet{Ivashina2011}, it can be used as a
generic noise model for any antenna system (\citet{Maaskant2010}). It is clear, however, that depending on the antenna and LNA designs, the electrical characteristics of the array and the sparsity and density of 
the station elements layout, the strength of mutual coupling varies. Thus, the term $\mathbf{R}_{rec}$ may be replaced by uncorrelated receiver noise in this definition. 

\subsection{Model-based calibration method}
\label{MBMeth.}

The model based calibration problem has been formulated as a least squares minimization problem by \citet{2009ITSP...57.3512W}. Here, we rewrite it with the refined data model given in Eq.\ref{eq:DM6}:

\begin{equation}
\{\hat{\mathbf{g}},\hat{\boldsymbol{\sigma}}_{\mathrm{n}}\}=\mathrm{argmin}_{\mathbf{g},\boldsymbol{\sigma}_n}\|\mathbf{G}\mathbf{M}\mathbf{A}\mathbf{\Sigma}_{\mathrm{s}}\mathbf{A}^H\mathbf{M}^H\mathbf{G}^H + \mathbf{\Sigma}'_{\mathrm{n}}-\hat{\mathbf{R}}\|_{F}^{2}
\label{eq:MSM1}
\end{equation}

This estimates the noise and complex gain of each receiver element using the measured visibility, $\hat{\mathbf{R}}$ and the modeled visibility, 
$\mathbf{G}\mathbf{M}\mathbf{A}\mathbf{\Sigma}_{\mathrm{s}}\mathbf{A}^H\mathbf{M}^H\mathbf{G}^H + \mathbf{\Sigma}'_{\mathrm{n}}$. $\mathbf{\Sigma}_{\mathrm{s}}$ and $\mathbf{A}$ are assumed to be known. 
We can calculate them, if we specify the time of observation, the telescope geometry and known source parameters.
In the presence of the coupling effects ($\mathbf{M}$ and $\mathbf{\Sigma}'_{\mathrm{n}}$), estimation results are biased, unless an accurate beam model of each individual element 
is provided or the matrix $\mathbf{M}$ is known. In that case, one can handle the effect of the correlated noise using a WALS approach (\citet{2010arXiv1003.2497W}).

Fig. \ref{Fig1} shows a sky map scanned by HBA tiles. One can see the Galactic plane due to the many short baselines and the sun as the dominant radio source. Given the beam models of the station elements, the WALS method treats the extended structure as correlated noise and estimates the parameters of interest. However, the model based methods are in general iterative methods which usually converge after several iterations.

\subsection{Redundancy calibration method}
\label{RedunMeth}

The basic assumption of redundancy calibration is that the redundant baselines sample the same Fourier component of the sky ($\mathbf{A}\mathbf{\Sigma}_{\mathrm{s}}\mathbf{A}^H$). This assumption holds if the array response vectors of the redundant baselines are the same i.e. the element beams are identical. We therefore begin with the most general data model given in Eq. \ref{eq:DM6} to understand the limitations of this method for a phased array station. 

To build up the system of equations for the redundancy calibration algorithm, we represent an off-diagonal element of $\mathbf{R}$ in Eq. \ref{eq:DM6} as:

\begin{equation}
R_{ij} = g_i g_j^{*} \left[\mathbf{M}\mathbf{A}\mathbf{\Sigma}_{\mathrm{s}}\mathbf{A}^H\mathbf{M}^H \right]_{ij} + \Sigma'_{\mathrm{n}_{ij}}
\label{eq:Redun1}
\end{equation}

Note that correlator offsets could contribute as additive corrupting factors in $\Sigma'_{\mathrm{n}_{ij}}$. Since $\mathbf{\Sigma}_{\mathrm{s}}$ in Sect. \ref{DaMod.} is a diagonal matrix, one can expand Eq. \ref{eq:Redun1} to:

\begin{align}
\label{eq:Redun2}
R_{ij} &= g_i g_j^{*} M_{ii} M_{jj}^* \overbrace{\sum_{q=1}^{q} A_{iq}\sigma_q A_{jq}^*}^\text{true visibility} \\
       &+ \underbrace{g_i g_j^{*} \sum_{q=1}^{q} \left(\sum_{\substack{p_1=1\\ p_1\neq i}}^{p} \sum_{\substack{p_2=1\\ p_2\neq
j}}^{p}M_{ip_1}M_{jp_2}^*A_{p_1q}A_{p_2q}^*\sigma_q\right)}_\text{additive term due to mutual coupling} \nonumber + \Sigma'_{\mathrm{n}_{ij}} 
\end{align}

This shows that the presence of the mutual coupling produces baseline dependent multiplicative and additive terms by influencing signals directly and noise indirectly ($\mathbf{R}_{rec}$ in Eq. \ref{eq:DM5}). 
This violates the fundamental assumption of redundancy i.e. we will not observe redundant visibilities on physically redundant baselines (see Fig. \ref{RedunnonRedun}, top).

Without loss of generality, $M_{ii} M_{jj}^* = 1$ (or they can be absorbed in the gains). To establish an analogy between Eq. \ref{eq:Redun2} and the redundancy method formalism in \citet{Wieringa1991}, we define the following term: 

\begin{equation}
e_{ij} = g_i g_j^{*} \sum_{q=1}^{q} \left(\sum_{\substack{p_1=1\\ p_1\neq i}}^{p} \sum_{\substack{p_2=1\\ p_2\neq j}}^{p} M_{ip_1} M_{jp_2}^*A_{p_1q}A_{p_2q}^*\sigma_q \right) + \Sigma'_{\mathrm{n},{ij}} 
\label{eq:Redun2_1}
\end{equation}

\noindent and rewrite the Eq. \ref{eq:Redun2} as:

\begin{equation}
{R}_{ij}^{obs} = {R}_{ij}^{true} g_i g_j^{*} + e_{ij}
\label{eq:Redun3}
\end{equation}

\noindent or 

\begin{equation}
R_{ij}^{obs} = {R}_{ij}^{true} {g}_i {g}_j^{*} \underbrace{(1 + \frac {e_{ij}} {{R}_{ij}^{true} {g}_i {g}_j^{*}})}_{w_{ij}}\\
\label{eq:Redun4}
\end{equation}

\noindent where ${R}_{ij}^{obs}$ and ${R}_{ij}^{true}$ are the observed and the theoretical redundant visibilities, respectively. $g_i$ and $g_j$ are the element complex gains. ${w}_{ij}$ can be defined as a baseline dependent error which affects the accuracy of the calibration results. 

We take the natural logarithm of both sides of the equation Eq. \ref{eq:Redun4} to obtain:

\begin{equation}
\mbox{ln}\ {R}_{ij}^{obs} =\mbox{ln} \arrowvert \gamma_i\arrowvert + \mbox{ln} \arrowvert \gamma_j \arrowvert + i(\phi_j - \phi_i) + \mbox{ln}\ {R}_{ij}^{true} + \mbox{ln} ({w}_{ij})\\
\label{eq:Redun5}
\end{equation}

In the absence of mutual coupling, the first term of $e_{ij}$ drops and the second term drops the contribution of $R_{rec}$. The only correlated contributing terms come from $R_{sky}$ and $R_{sp}$ which are negligible depending on the SNR of the observation. Lets assume that we have such a case where we can ignore $e_{ij}$. Then, we equate the amplitude and the phase values to decouple the system of equations for the phases and the amplitudes:

\begin{equation}
\mbox{ln} \arrowvert{R}_{ij}^{obs}\arrowvert = {\gamma}_i^{'} + {\gamma}_j^{'} + \mbox{ln} \arrowvert{R}_{ij}^{true}\arrowvert 
\label{eq:Redun6}
\end{equation}

\begin{equation}
{\psi}_{ij}^{obs} = {\phi}_j - {\phi}_i + {\psi}_{ij}^{true} 
\label{eq:Redun7}
\end{equation}

\noindent $\arrowvert{R}_{ij}\arrowvert$ and ${\psi}_{ij}$ are the amplitude and the phase of a complex visibility. Since we have to specify the absolute flux level, we set:

\begin{equation}
\Sigma \gamma_i^{'} = 0
\label{eq:Redun8}
\end{equation}

\noindent We also have to constrain the element phase. We can enforce this constraint by specifying that the average phase for all elements is zero:

\begin{equation}
\Sigma \phi_i = 0
\label{eq:Redun9}
\end{equation}

\noindent Furthermore, there might also be an arbitrary linear phase slope over the array. This phase slope corresponds to a position shift of the field. This arises because a redundancy solution does not provide an absolute position. This can either be absorbed in the true visibilities or in the element phases. For a two dimensional array, we constrain x and y in the same manner: 

\begin{equation}
\Sigma_{i=1}^p \phi_i x_i = 0
\label{eq:Redun10}
\end{equation}

\begin{equation}
\Sigma_{i=1}^p \phi_i y_i = 0
\label{eq:Redun11}
\end{equation}

\noindent where $x_i$ and $y_i$ are the $(x, y)$ coordinates of the array elements. Eq. \ref{eq:Redun6}- Eq. \ref{eq:Redun11} formulate the redundancy calibration method as two overdetermined systems of linear equations for phases and amplitudes which can be solved in a single step least square solution. For instance, the phase estimator can be symbolized as $\mathbf{E}_{ph}\boldsymbol{\theta} = \mathbf{\Psi}^{obs}$ and solved using the pseudo-inverse:

\begin{equation}
\hat{\boldsymbol{\theta}} = \left[\mathbf{E}_{ph}^T \mathbf{E}_{ph}\right]^{-1} \mathbf{E}_{ph}^T \mathbf{\Psi}^{obs}
\label{eq:Redun12}
\end{equation}

\noindent where $\boldsymbol{\theta} = [\phi_1, \phi_2, ..., \phi_p, {\psi}_{1}^{true}, {\psi}_{2}^{true}, ..., {\psi}_{m}^{true}]$ is the vector of parameters to be estimated, $m$ is the number of distinct 
redundant baselines, $\mathbf{E}_{ph}$ is the coefficient matrix and $\mathbf{\Psi}^{obs}$ is the vector of the observed redundant phases and right sides of Eq.\ref{eq:Redun9}- Eq.\ref{eq:Redun11}. Setting a
 phase reference for element phases, is done after the phase estimation.

Note that we came to the solution in Eq. \ref{eq:Redun12}, because we ignored the baseline dependent noise. In its presence, our problem gets the form of $\mathbf{E}_{ph}\boldsymbol{\theta} + \boldsymbol{\beta}= \boldsymbol{\Psi}^{obs}$ (expanded in Eq. \ref{eq:Redun14}). Then, the estimated parameters will deviate as follows:

\begin{equation}
\boldsymbol{\varepsilon} \equiv {\boldsymbol{\theta} - \hat{\boldsymbol{\theta}}} = \left[\mathbf{E}_{ph}^T \mathbf{E}_{ph}\right]^{-1} \mathbf{E}_{ph}^T \boldsymbol{\beta}
\label{eq:Redun13}
\end{equation}

\noindent where $\boldsymbol{\beta} = \mathrm{vec}(\angle(\mbox{ln} ({w}_{ij})))$. If the array mutual coupling is significant, the redundancy method will not be a reliable estimator. 
In the case of a weakly coupled array, the vector $\boldsymbol{\beta}$ still carries the correlated noise due to ${R}_{sp}$ and ${R}_{sky}$. Assuming that these have Wishart distribution in Eq.\ref{eq:Redun2_1},
their statistical distribution changes in vector $\boldsymbol{\beta}$. This has been taken into account for the results in Sect. \ref{CalQ.}.

\begin{figure*}
\begin{center}
\begin{equation}
\underbrace{\left[
\begin{array}{ccccccccccccccccccc}
1 & -1 & 0 &  0 & 0  & 0 & 0 & 0 & 0 & 0 & 0 &... & 0 & 0 & 1 & 0 & 0 &...& 0\\
0 &  0 & 1 & -1 & 0  & 0 & 0 & 0 & 0 & 0 & 0 &... & 0 & 0 & 1 & 0 & 0 &...& 0\\
0 &  0 & 0 &  1 & -1 & 0 & 0 & 0 & 0 & 0 & 0 &... & 0 & 0 & 1 & 0 & 0 &...& 0\\
0 &  0 & 0 &  0 &  1 & -1& 0 & 0 & 0 & 0 & 0 &... & 0 & 0 & 1 & 0 & 0 &...& 0\\ 
. &    &   &    &    & . &   &   &   &   &   &    &   &   &   &   &.  &   & 0\\
. &    &   &    &    & . &   &   &   &   &   &    &   &   &   &   &.  &   & 0\\
. &    &   &    &    & . &   &   &   &   &   &    &   &   &   &   &.  &   & 0\\
0 &  0 & 0 &  0 &  0 & 0 & 0 & 0 & 0 & 0 & 0 &... & 1 &-1 & 1 & 0 & 0 &...& 0\\
. &    &   &    &    & . &   &   &   &   &   &    &   &   &   &   &.  &   & 0\\
. &    &   &    &    & . &   &   &   &   &   &    &   &   &   &   &.  &   & 0\\
. &    &   &    &    & . &   &   &   &   &   &    &   &   &   &   &.  &   & 0\\ 
1 & 0  & 0 &  0 & 0  & 0 & 0 & 0 & 0 & 0 & 0 &... & -1& 0 & 0 & 0 & 0 &...& 1 \\
0 & 1  & 0 &  0 & 0  & 0 & 0 & 0 & 0 & 0 & 0 &... & 0 &-1 & 0 & 0 & 0 &...& 1 \\
1 & 1  & 1 &  1 & 1  & 1 & 1 & 1 & 1 & 1 & 1 &... & 1 & 1 & 0 & 0 & 0 &...& 0 \\
x_1 & x_2 & x_3 & x_4 & . & . & . & . & . & . & . &... & . & x_{24} & 0 & 0 & 0 &...& 0 \\
y_1 & y_2 & y_3 & y_4 & . & . & . & . & . & . & . &... & . & y_{24} & 0 & 0 & 0 &...& 0 \\
\end{array} 
\right]}_{\mathbf{E}_{ph}} \underbrace{\left[
\begin{array}{c}
\phi_1\\
\phi_2\\
\phi_3\\
.\\
.\\
.\\
\phi_{24}\\
{\psi}_{1}^{true}\\
{\psi}_{2}^{true}\\
{\psi}_{3}^{true}\\
.\\
.\\
.\\
{\psi}_{36}^{true}\\
\end{array}
\right]}_{\boldsymbol{\theta}} + \underbrace{\left[
\begin{array}{c} 
{\beta}_{1,2}\\
{\beta}_{3,4}\\
{\beta}_{4,5}\\
{\beta}_{5,6}\\
.\\
.\\
.\\
{\beta}_{23,24}\\
.\\
.\\
.\\
{\beta}_{1,23}\\
{\beta}_{2,24}\\
0\\
0\\
0\\
\end{array}
\right]}_{\boldsymbol{\beta}} = \underbrace{\left[
\begin{array}{c} 
{\psi}_{1,2}^{obs}\\
{\psi}_{3,4}^{obs}\\
{\psi}_{4,5}^{obs}\\
{\psi}_{5,6}^{obs}\\
.\\
.\\
.\\
{\psi}_{23,24}^{obs}\\
.\\
.\\
.\\
{\psi}_{1,23}^{obs}\\
{\psi}_{2,24}^{obs}\\
0\\
0\\
0\\
\end{array}
\right]}_{\boldsymbol{\Psi}^{obs}}
\label{eq:Redun14}
\end{equation}
\end{center}
\end{figure*}

Since the created systems of equations are highly sparse, they are computationally fast. Most importantly, they are independent of a sky model but their accuracy is affected by the SNR of the observed sky as discussed by \citet{2010MNRAS.408.1029L}. They require identical beams of station elements and a regular arrangement of antennas to provide a sufficient number of redundant baselines. Then, one can solve for the element complex gains in a single step estimation. 

\section{Implementation of the redundancy calibration using real and simulated HBA data}
\label{Impl.}

\subsection{Verification of the fundamental assumption of redundancy}
\label{RedunChek.}

\begin{figure*}[!ht]
\centering
\includegraphics [width=16cm] {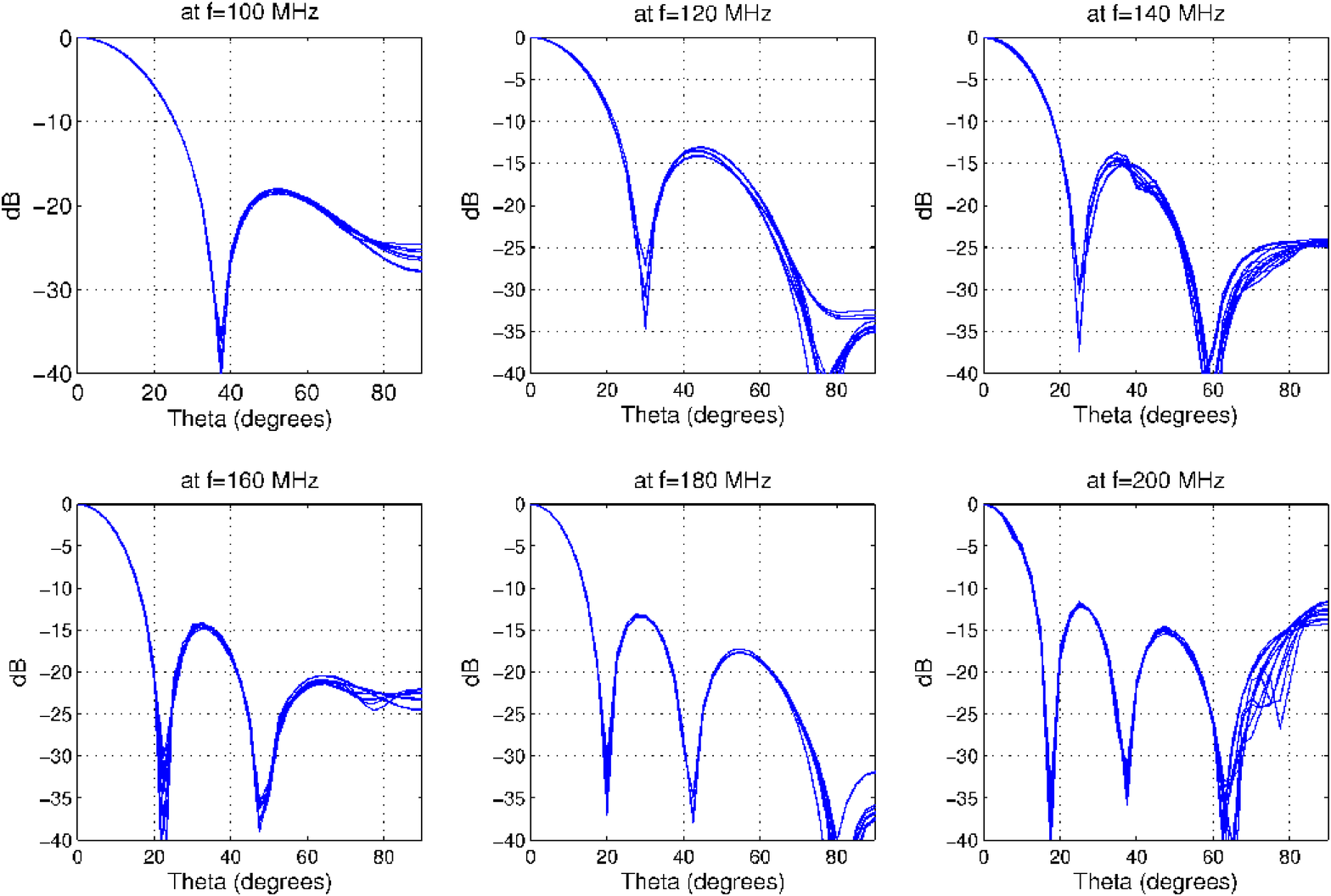}
\caption{Simulated beam pattern of the tiles in a 24-tile HBA station using CAESAR. Each subplot presents the beam patterns of 24 tiles in a particular frequency in $\phi = 0$ plane. The beams are formed toward the local zenith. The non-identicalness in the sidelobes is caused by the different mutual coupling environments of each tile.}
\label{Fig3}
\end{figure*}

\begin{figure*}[!ht]
\centering
\includegraphics [width=16cm] {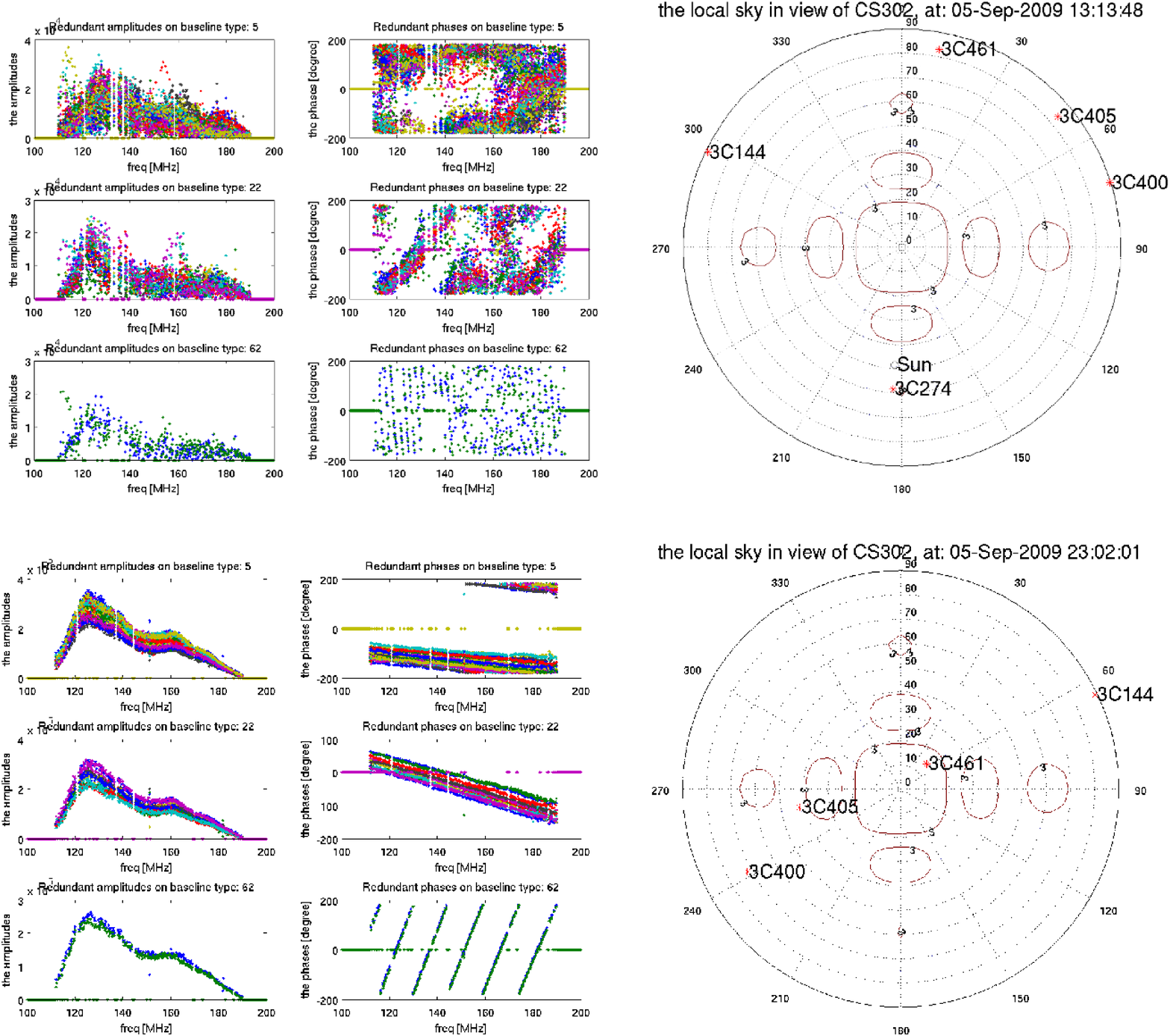}
\caption{An observation with a HBA station on September 5, 2009 at 13:13:48 UTC in which redundancy calibration fails (top) and an observation with the same sation at 23:02:01 UTC in which redundancy 
calibration is successful (bottom). The left panels show the measured visibilities for three distinct type of redundant baselines. The right panels show the local sky model at the time of observation with 
strongest sources on the sky superimposed on a contour plot of the element beam pattern at 170 MHz.}
\label{RedunnonRedun}
\end{figure*}

As mentioned earlier, redundancy translates to checking the identicalness of the element beams in a station. We have checked this for a 24-tile HBA station like the one shown in Fig.\ref {Fig2} using CAESAR (Computationally Advanced and Efficient Simulator for ARrays) (\citet{Maaskant2008}; \citet{2006ESASP.626E.389M}). Numerical computation for the EM simulation was done using the CBFM (Characteristic Basis Function Method) which is the most numerically efficient and accurate method available for large scattering problems (\citet{Prakash2003}; \citet{Yeo2003}). Fig. \ref{Fig3} presents the simulation results when the beams are formed toward the local zenith. Each subplot presents the beam patterns of 24 tiles for a particular frequency in the $\phi = 0$ plane. One can see that due to a different mutual coupling environment, each tile has a slightly different beam pattern. The mutual coupling effect is sufficiently small not to disturb the main beam but only the far sidelobes which are at least $12 \mathrm{dB}$ lower than the main lobe. However, it is not a favorable condition for redundancy calibration in general. Based on this simulation, 
we expect to observe non-redundant visibilities on redundant baselines when a strong source falls in the sidelobes. This has been confirmed by the real observation presented in Fig. \ref{RedunnonRedun}, top. In this observation, due to the absence of a strong source in the main beam, the non-identical sidelobes have the chance to play a significant role in disturbing the redundancy while in the observation shown in Fig. \ref{RedunnonRedun}, bottom, one strong source in the main beam seems to be sufficient to dominate the influence of other possible strong sources which 
are observed through the non-identical sidelobes. There is of course a different contribution from the correlated background noise or sky noise ($\mathbf{R}_{sky}$) in Fig. \ref{RedunnonRedun}, top and bottom panels.
Fig. \ref{RedunnonRedun} presents two data sets obtained to verify the use of redundancy in a HBA station. The observations were made on 5 September 2009.
The tile beams in a HBA station (in the LOFAR core, known as CS302) were formed toward the local zenith. We let the sky drift over the FoV (Field of View) of the station. We captured the station visibilities 
approximately every 10 minutes (integration time of one second per frequency channel for 512 frequency channels). The bandwidth of the frequency channels is $195$ kHz and
the frequency range is $100-200$ MHz. On the right panel, the local sky viewed from CS302 is presentedn to show which sources have contributed to the visibilities shown on the left panel.

Since, the mutual coupling environment changes at lower elevations, the identicalness of the main beams is disturbed consequently up to $1-2 \mathrm{dB}$. Real observations have shown less disturbance in the main beam at lower elevations (see Sect. \ref{CalQ.} and Appendix. \ref{appendix:a}).

\begin{figure*}[!ht]
\centering
\includegraphics [width=12cm] {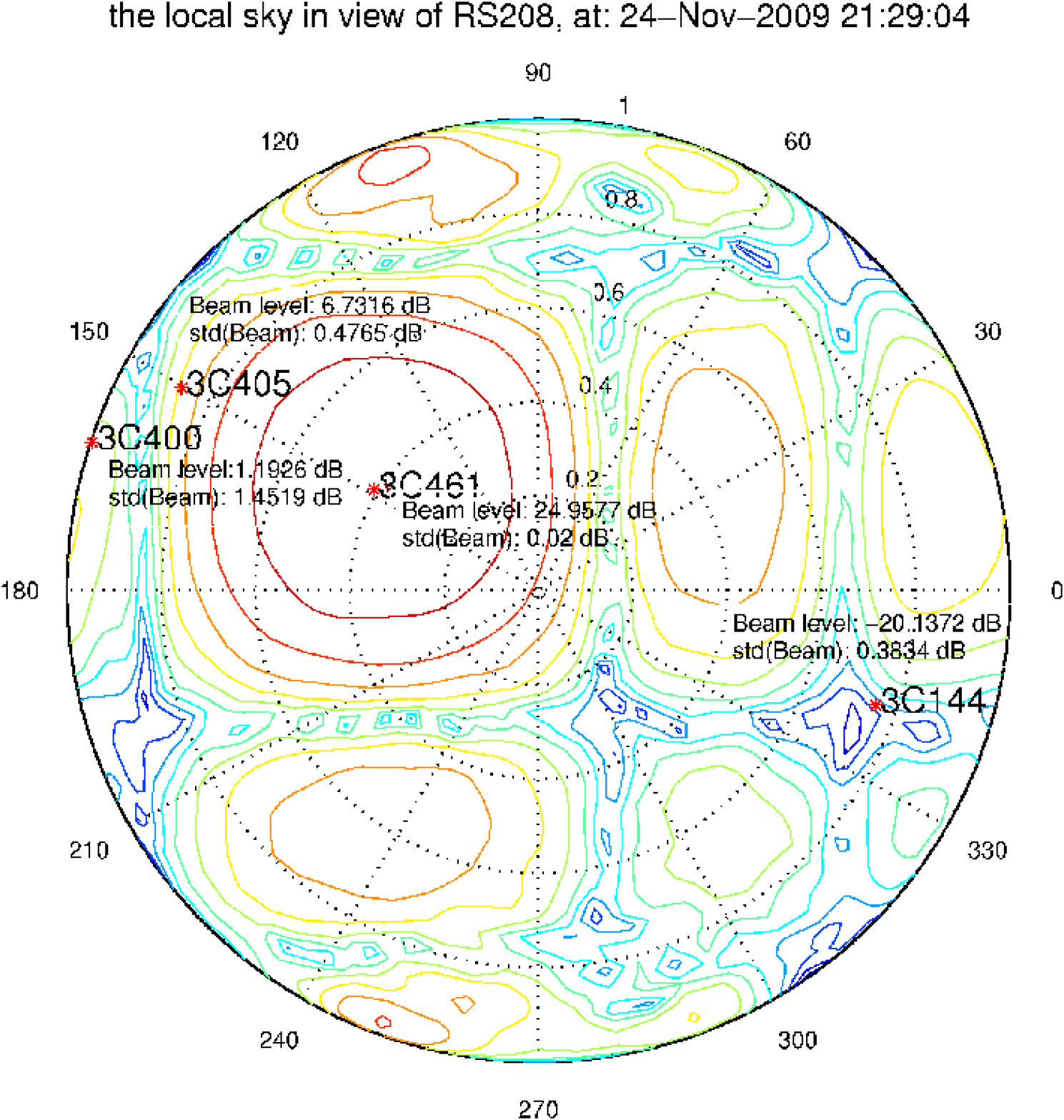}
\caption{A-team radio sources in the FoV of an HBA station called RS208 on 24th November 2009 at 21:29:04 UTC. Tile beams are streered toward 3C461 to provide the most redundant visibilities. 
The values of the averaged reception pattern of 48 tiles, $A(\sigma)$ and their standard deviation, $std(A(\sigma))$ at $120$MHz have also been indicated at other sources locations, $\sigma$.}
\label{Fig6}
\end{figure*}

\begin{figure*}[!ht]
\centering
\includegraphics [width=16cm] {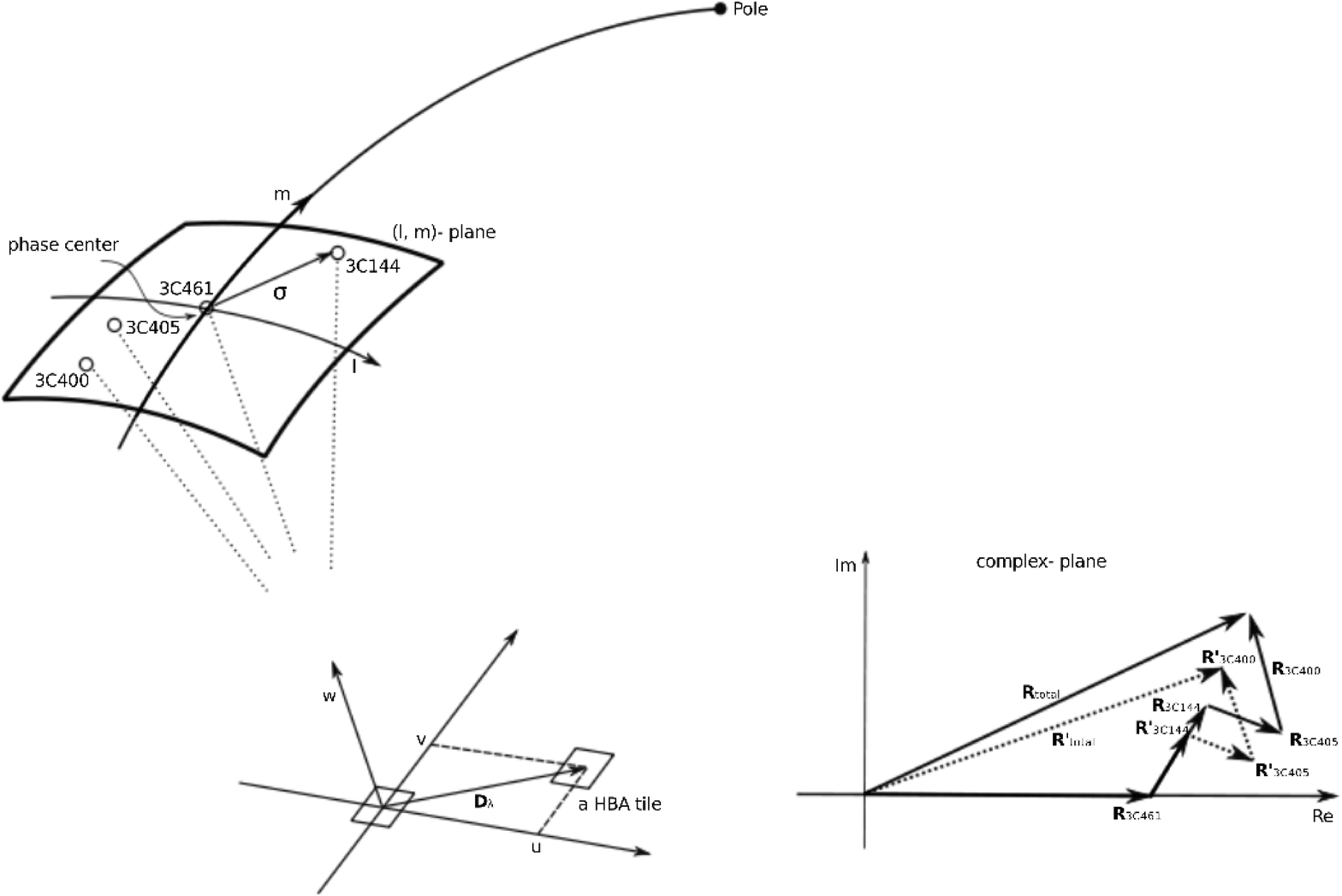}
\caption{Total complex visibility on a certaion baseline due to the contribution of all strong sources shown in Fig.\ref{Fig6}. The ones which contribute through non-identical sidelobes introduce non-redundancy or systematic errors to the total redundant visibility. This has been demostrated in a complex plane on the right. The solid line shows the visibility vectors as they were observed through hypothetically identical sidelobes. The dotted line shows the visibility vectors when they are attenuated differently by the actual non-identical sidelobes.}
\label{Fig7}
\end{figure*}

Based on the EM simulation by CAESAR (Fig. \ref{Fig3}) and real observations (Fig. \ref{RedunnonRedun}), we conclude that the best case scenario for redundancy calibration of an HBA station is to have a strong
source in the main lobe when the beams are identical. However, the non-identicalness of the sidelobes introduces non-redundancy or systematic errors which can not be eliminated by any statistical method or
longer integration time. Therefore, we must investigate whether the contribution of other strong sources, which may fall in the non-identical sidelobes, is significant. To quantify this systematic error on
the visibility measured on a given baseline, we present the following example. At 21:29:04 UTC on 24th November 2009, at an HBA station called RS208 located at ($lon= 6.9196^{\circ} lat= 52.6699^{\circ}$),
four strong sources are in the FoV as shown in Fig. \ref{Fig6}. In Fig. \ref{Fig6}, the tile beams have been formed toward 3C461 or CasA. Thus, it is in the phase center.
The sources 3C405 (CygA), 3C400 and 3C144 (TauA) have fallen in the sidelobes. The tile beams and their standard deviation at $120$MHz, at these sources are also depicted in Fig. \ref{Fig6}. 
The total complex visibility observed on a certain baseline, $\mathbf{D}_{\lambda}$ is computed as:

\begin{equation}
\mathbf{R}^{obs} = \arrowvert{R^{obs}}\arrowvert e^{j\psi^{obs}} = \int_{s} \varLambda(\boldsymbol{\sigma}) I(\boldsymbol{\sigma}) e^{-j2\pi \mathbf{D}_{\lambda}.\boldsymbol{\sigma}} d\Omega
\label{eq:TotVis}
\end{equation}

\noindent where $\varLambda(\boldsymbol{\sigma}) \equiv A(\boldsymbol{\sigma})/A_0$ is the normalized tile reception pattern at $\sigma$, $A_0$ being the response at the beam center.
$I(\boldsymbol{\sigma})$ is the source flux. Since there are four dominant strong sources in the FoV, we assume that the integral can be replaced by a summation. 
The complex plane in Fig. \ref{Fig7} right, shows summations of the visibility vectors as they were observed through hypothetically identical sidelobes, $\mathbf{R}_{total}$ (solid line) as well as their 
summation when they are attenuated differently by the actual non-identical sidelobes, $\mathbf{R}^{'}_{total}$ (dotted line). We compute the systematic errors in the phase and the amplitude of the 
visibility as follows: we compute the complex vector, $\mathbf{R}_{total} = \mathbf{R}_{3C461} + \mathbf{R}_{3C144} + \mathbf{R}_{3C405}+ \mathbf{R}_{3C400}$ using Eq. \ref{eq:TotVis}. For this,
we choose the baseline type
given in Fig. \ref{Fig9}, right most panel ($\mathbf{D}$) and frequency, $f= 120$MHz for $\mathbf{D}_{\lambda} = \mathbf{D}/\lambda$, $\lambda$ is the wavelength. We set the averaged beam pattern of all 48
tiles as the identical beam for the term $A(\boldsymbol{\sigma})$ whose values are shown in Fig. \ref{Fig6} at different source locations, as the quantity beam level.
We use $\varLambda(\boldsymbol{\sigma}) = A(\boldsymbol{\sigma})/ A(\boldsymbol{\sigma}_{3C461})$. The source fluxes, $I(\boldsymbol{\sigma})$ are given in Table. \ref{tab:cal_source}. 
We also compute $\mathbf{R}^{'}_{total} = \mathbf{R}^{'}_{3C461} + \mathbf{R}^{'}_{3C144} + \mathbf{R}^{'}_{3C405}+ \mathbf{R}^{'}_{3C400}$. The computation settings are as before except that 
$A(\boldsymbol{\sigma})$ is not identical this time. It is deviated from the averaged beam pattern by the standard deviation values shown in Fig. \ref{Fig6} at different source locations, as 
the quantity $std(Beam)$. The bias introduced in the amplitude and the phase of $\mathbf{R}^{'}_{total}$ as compared with $\mathbf{R}_{total}$ are presented in the first row of Table. \ref{tab:cal_source}.

We steer the beam toward each availabe strong source and repeat the calculation to predict the systematic error due to the other ones in the sidelobes with different
standard deviations. The results are presented in Table. \ref{tab:cal_source}. We can conclude that tracking CasA provides the best redundancy in the observed visibilities.

\begin{table*}[!ht]
\caption{A-team radio sources in the FoV of RS208, on 24th of November 2009 at 21:29:04 UTC and the predicted levels of the systematic errors in the amplitudes and the phases of the redundant visibility
 due to the contribution of other strong sources through non-identical sidelobes. The baselines given in Fig. \ref{Fig9}, right most panel and frequency, $f= 120$MHz were chosen for the computations.} 
\centering 
\renewcommand{\footnoterule}{} 
\begin{tabular}{l llll} 
\hline\hline 
Calibrator source& source flux $[Jy]$& normalized error in amplitudes& error in phases $[rad]$\\[0.5ex]
\hline 
3C461 (CasA) & $8609$ &$0.07$ & $ 0.06$\\[1ex] 
3C405 (CygA) & $8100$ &$0.20$ & $ 0.09$\\[1ex] 
3C400 & $540$ & $0.93$ &$ 0.01$\\[1ex] 
3C144 (TauA) & $1420$ &$0.84$ & $ 0.28$\\[1ex]
\hline
\end{tabular} 
\label{tab:cal_source}
\end{table*}

The plot of the residuals for the corrected redundant visibilities shown in Fig. \ref{Fig9} confirms the results in Table. \ref{tab:cal_source}. Fig. \ref{Fig9} reveals non-redundancy in the measured visibilities (in terms of amplitudes and phases). In Table. \ref{tab:cal_source}, we predict the same quantities which are caused by non-identical sidelobes. There is a few percent of discrepancy between the predicted residuals and their actual values in our observation. This can be explained by our simplifying assumption that we made for the number of strong sources in the FoV. However, Table. \ref{tab:cal_source} can play an instrumental role for station calibration. 

\subsection{Redundancy calibration performance}
\label{CalQ.}

\begin{figure*}[!ht]
\centering
\includegraphics [width=16cm] {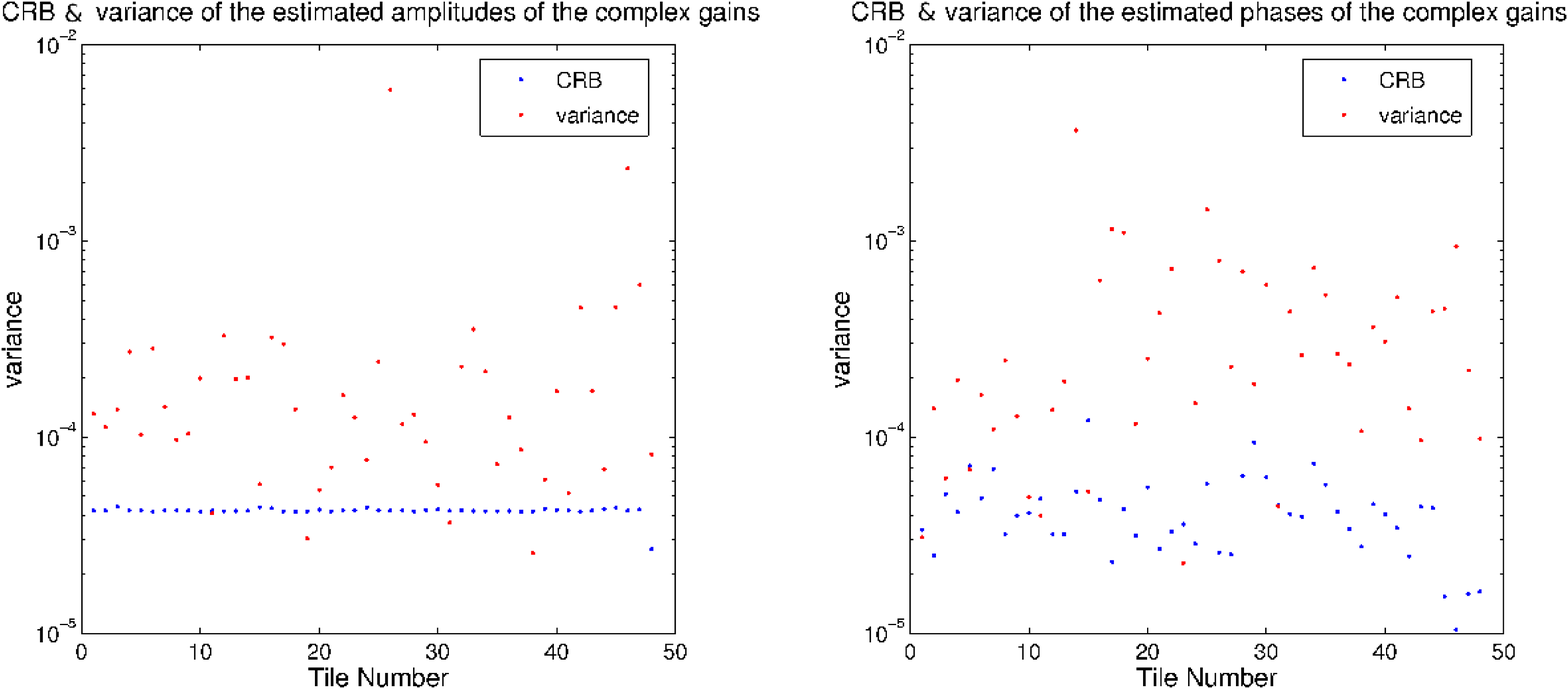}
\caption{Theoretical minimum variance of the estimated amplitudes and phases of the complex gains or CRB compared with their actual variance. These quantities for the amplitudes and the phases are presented in two separate plots as their estimators were decoupled in Sect. \ref{RedunMeth}.}
\label{Fig8}
\end{figure*}

To study the redundancy calibration performance, we tracked CasA on 24th November 2009 from 15:25:43 UTC until 22:12:19 UTC. Running redundancy calibration on 48 captured data sets during this observation, 
gives very stable results for the receiver complex gains over time. This is an indication of system stability and a working calibration routine, which was running approximately every 10 minutes. We compare
the variance of the estimated complex gains over 48 runs of redundancy calibration with the CRB in Fig. \ref{Fig8}. These quantities for the amplitudes and the phases are presented in two separate plots as
their estimators were decoupled in Sect. \ref{RedunMeth}. The CRB (or the theoretical minimum variance) and the actual variance on the estimated parameters over time are in good agreement. The small difference between them can be explained by not having exactly the same SNR from one observation to another during our survey from 15:25:43 UTC until 22:12:19 UTC while the predicted SNR for this observation is SNR $ \simeq 0.75$, following the analysis by \citet{SJWijnholds2011}. This value was used to compute the presented CRB in Fig. \ref{Fig8}. 

\begin{figure*}[!ht]
\centering
\includegraphics [width=17cm] {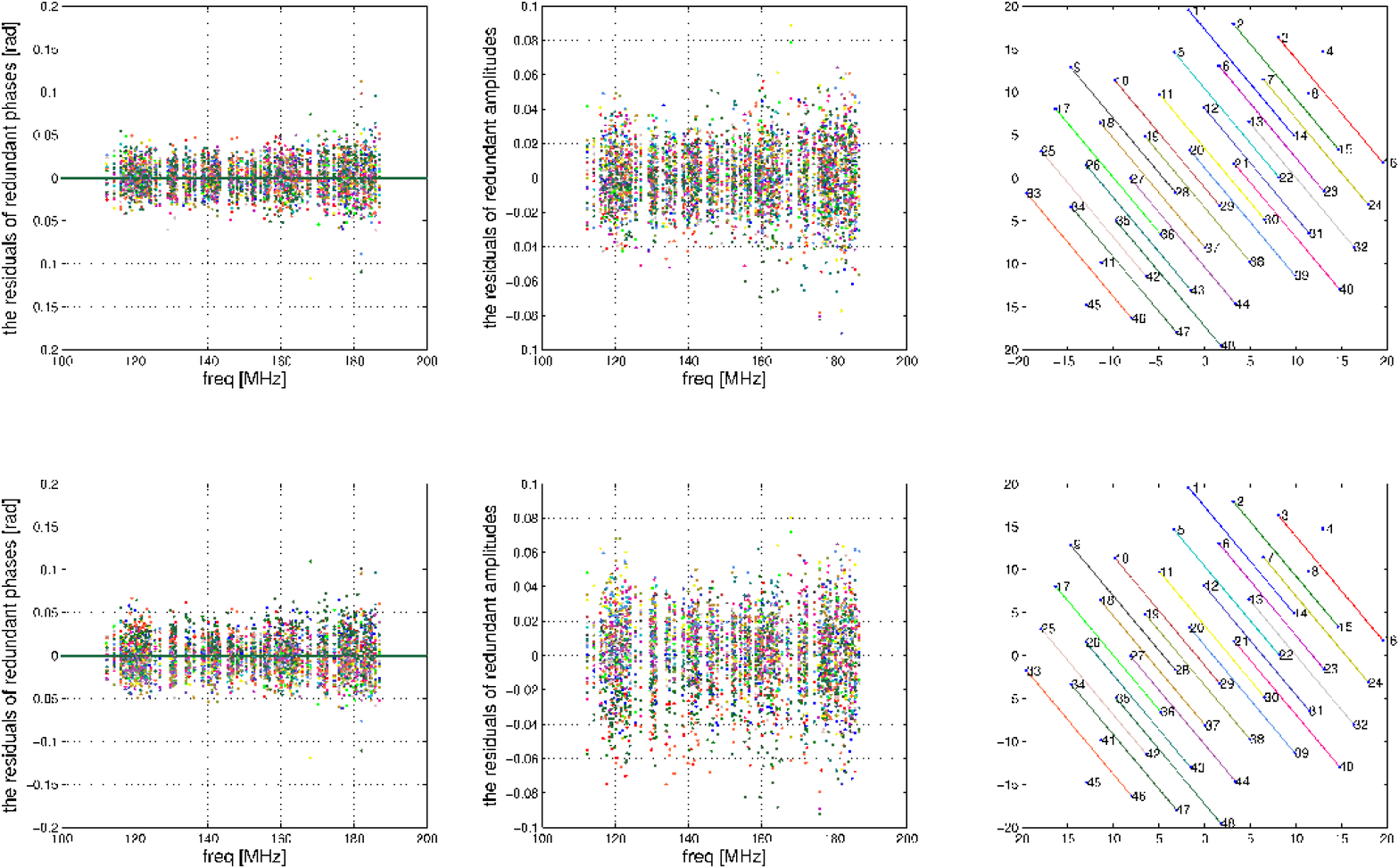}
\caption{Plots of the residuals for corrected redundant visibilities in terms of phases and amplitudes on a given set of redundant baselines. The first row shows the results after redundancy calibration. The second row shows the result afters model based calibration. The data is taken from the observation done on 24th November 2009 at 21:29:04 UTC, when CasA is at high elevation. The station configuration of RS208 is shown in the right most panel. The corresponding redundant baselines to each redundant visibility are also, with the same color code depicted on the station layout.}
\label{Fig9}
\end{figure*}

We also studied the residuals for the corrected redundant visibilities. Fig. \ref{Fig9} shows an example of the residuals in amplitudes and phases of the corrected visibilities on a distinct type of 
redundant baseline. The snapshot is captured at 21:29:04 UTC, when CasA is at high elevation. The first row shows the results after redundancy calibration as compared to the results after the model based
calibration in the second row. The type of redundant baseline is depicted on the station configuration in the right most panel. The integration time is 1 second per frequency channel. More plots of the residuals
from the same observation are presented in Appendix.\ref{appendix:a}, Fig. \ref{Fig_a1} and Fig. \ref{Fig_a2}. By comparing them, one may notice that the baseline length does not make a significant difference
in the residuals. The residuals are in the order of $2-5\%$ in both phases and amplitudes. Because a strong source like CasA in the field of view dominates the effect of possible correlated noise, this can be 
explained by not having $100\%$ identical main beams or the corruption due to the nonidentical sidelobes (see Table. \ref{tab:cal_source}). However these results are satisfactory. Moreover, Fig. \ref{Fig8} showed,
the estimated complex gains which are our parameters of interest in a station calibration, are very stable. This is due to the constraints on them in the formalism of the redundancy calibration.

Using the data captured at 15:25:43 UTC, when CasA is at low elevation, reveals slightly larger residuals in the order of $5-10\%$ (see Appendix.\ref{appendix:a}, Fig. \ref{Fig_a3}- Fig. \ref{Fig_a5}).
This is due to a different mutual coupling environment which leads to less identical beams either in the main lobe or the sidelobes.

The residuals also show that the two calibration methods perform almost equally well, although redundancy method behaves more consistently. Model based calibration has slightly larger residuals on short baselines,
e.g. up to $\sim0-1\%$ in phases and $\sim0-2\%$ in amplitudes whereas, it shows similar residuals as redundancy calibration on long baselines. Extended structures e.g. the galactic plane or north polar spur are captured on short baselines. 
Modeling them is computationally expensive and somehow impractical. Therefore, in the model based method, one has to discard the visibilities measured on short baselines to simplify the measured sky for a 
corresponding simple sky model while the redundancy method is sky model independent. Experiment has shown that discarding the equations of short baselines in redundancy calibration routine, will not improve its
results significantly.

Moreover, model based methods are sensitive to RFI sources, as their presence confuses the sky model. Since redundancy calibration is independent of the sky model, it is less sensitive to RFI. One may have noticed
that more frequency channels had to be flagged for model based calibration.

\begin{figure*}[!ht]
\centering
\includegraphics [width=16cm] {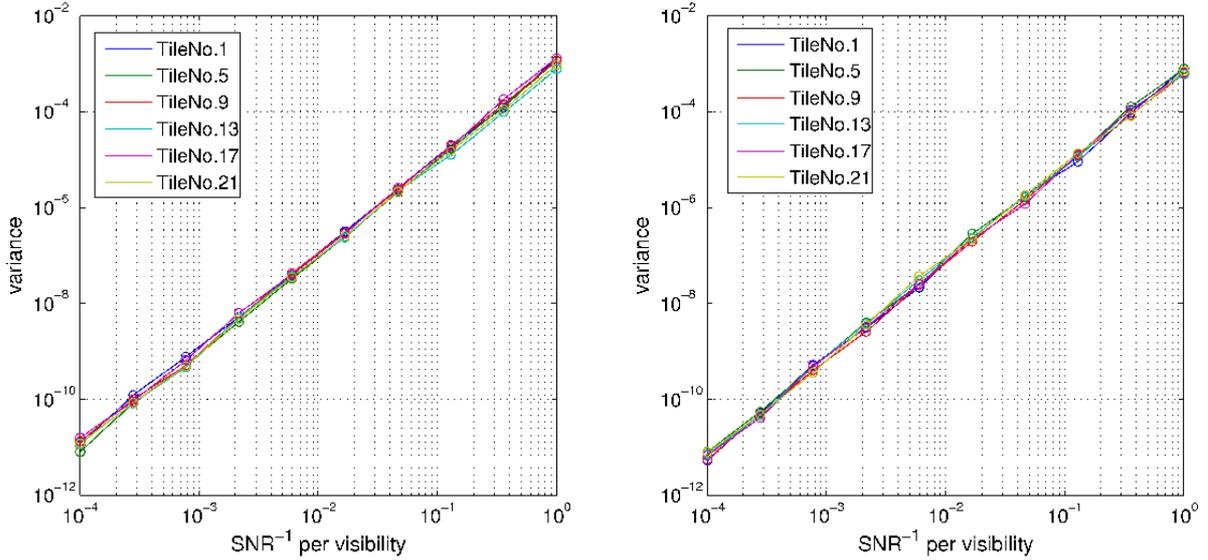}
\caption{The variance of the estimated complex gains in terms of amplitude (left) and phase (right) versus the $SNR^{-1}$ per visibility. Each color represents the estimated complex gain of a tile whose number is mentioned in the legend.}
\label{Fig12}
\end{figure*}

As discussed by \citet{2010MNRAS.408.1029L}, redundancy calibration quality depends on the SNR of the observed sky although it is independent of a sky model. \citet{2010MNRAS.408.1029L} showed that the estimated
parameters are affected differently in the presence of baseline dependent noise assuming it is Gaussian noise. The vector $\boldsymbol{\beta}$ in Eq. \ref{eq:Redun13} reveals a non-Gaussian baseline dependent noise.
We have taken this into consideration for the following results. After adding different levels of a non- i.i.d. Gaussian noise to the output vector of the array (Eq. \ref{eq:DM1}), 
we simulated the station visibility assuming that the station elements have identical beams. We used the station configuration of RS208 (see Fig. \ref{Fig9}, right most panel). 
The complex gains were estimated over 100 runs of Monte-Carlo simulation for different SNR per visibility. The variance of the error in the estimated complex gains (over simulation runs with different random
noise) versus different SNR per visibility is shown in Fig. \ref{Fig12}. These plots indicate the variance of the estimated complex gains which are our parameters of interest in a station calibration, 
are the same for the different tiles. This shows the stability of the algorithm. However the analysis like this is very array configuration dependent, as the array configuration determines the coeficient matrices by which our estimators are defined.

\subsection{Limitations of the redundancy calibration}
\label{RedunLim}

The applicability of redundancy calibration is limited by the following factors:

\begin{enumerate}
 \item The station configuration
 \item Mutual coupling between the station elements which leads to:\begin{itemize}
	\item non-identical element beams
 	\item presence of baseline dependent noise\end{itemize}
 \item SNR of the observed sky 
\end{enumerate}

\begin{figure*}[!ht]
\centering
\includegraphics [width=16cm] {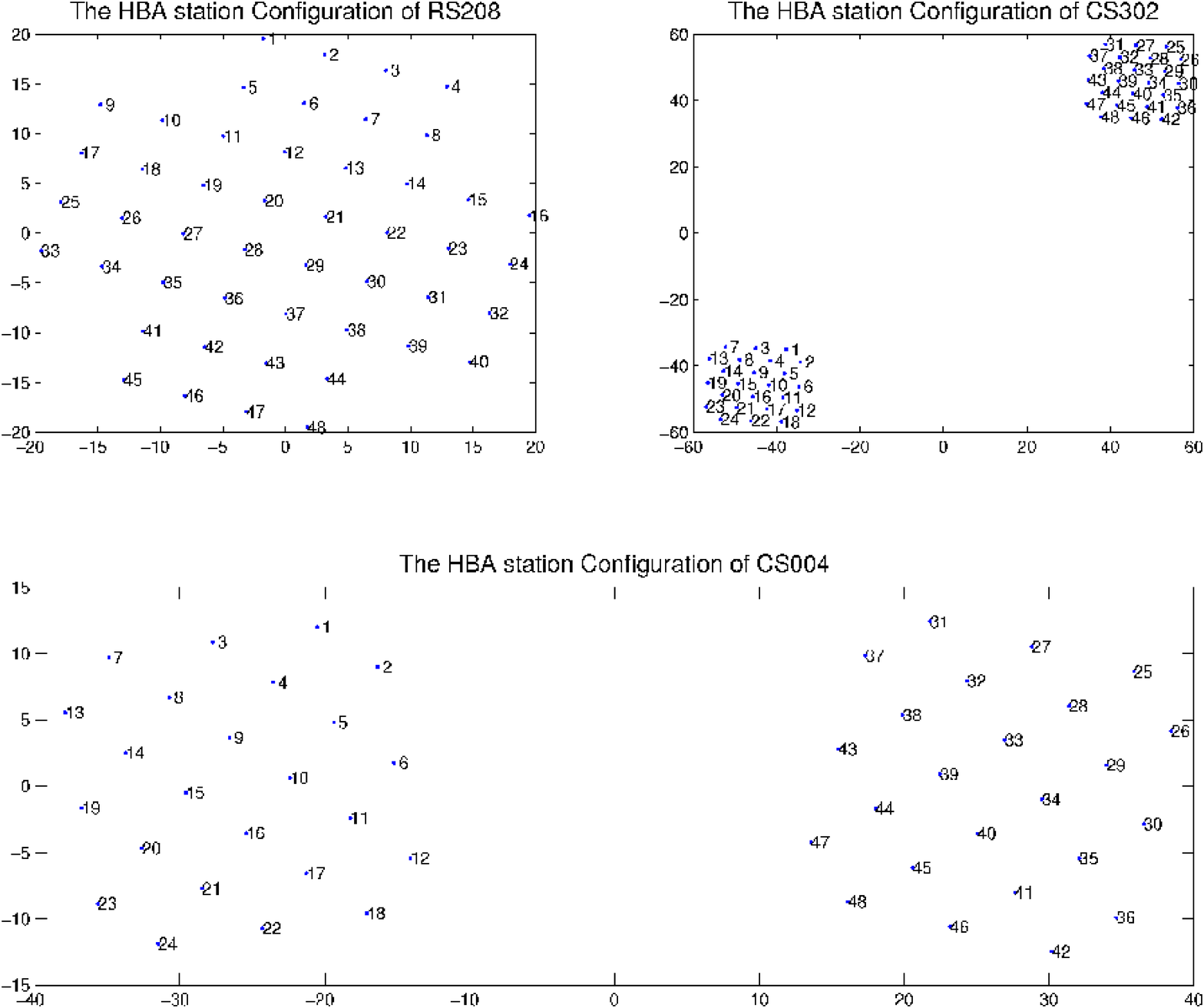}
\caption{Three different station configuration for HBA stations.}
\label{Fig14}
\end{figure*}

The station configuration can influence the suitability of redundancy in different ways. The redundancy calibration method requires a regular arrangement of station elements. In a station with $p$ elements one needs
a sufficient number of distinct types of redundant baseline to have a system of equations in which all station element gains are involved. The more redundant the station's baselines are, the less information
in the measured visibilities will be missed in the computation. This is not a concern in the HBA stations, as they are highly redundant. We define the ratio of the number of the measured visibilities that 
are used in the computations to the total number of the measured visibilities as $I$. The station configuration also determines the coeficient matrices i.e. $\mathbf{E}_{ph}$ or
$\mathbf{E}_{ampl}$ for phase and amplitude estimators respectively. The condition number is a relative error magnification factor i.e. errors in the right-hand side of a linear system of equations can cause
errors $\kappa(\mathbf{E}_{ph})$ times as large in the solution. Fig. \ref{Fig14} shows three different configurations for the HBA stations in the LOFAR system. Given a certain configuration, after 
recognizing the redundant baseline in it, one can calculate the aforementioned quantities. These are given in Table. \ref{tab:conf_cond}. This table shows that RS208 is the most reliable configuration, from the redundancy calibration point of view. However, these quantities can be figures of merit when we design for redundancy within the stations and in arrangement of the stations within the whole array.

Configuration and element spacing in a station are usually decided based on obtaining low sidelobes of the station beam. Minimizing the mutual coupling between the station elements should be taken into consideration as other important figures of merit in a station layout for the sake of redundancy applicability. This consideration serves a twofold goal, obtaining identical element beams at least in the main beam and minimizing the receiver correlated noise ($\mathbf{R}_{rec}$). Non-identicalness of the element beams introduces systematic errors which can not be eliminated by any statistical method or longer integration time. In the presence of the baseline dependent noise, the estimated parameters will deviate as given by Eq. \ref{eq:Redun13}. Quantifying this deviation for a given array, requires a good understanding of the noise terms in Eq. \ref{eq:DM5} especially $\mathbf{R}_{rec}$. The correlated noise terms are hard to model analytically, \citet{Maaskant2010}. One may use a powerful and generic tool called CAESAR to compute them numerically using its EM- (ElectroMagnetic) and MW- (MicroWave) simulators. This is an ongoing work for HBA and EMBRACE stations. Primary results from the MW-simulation for an HBA station has shown a negligible contribution of $\mathbf{R}_{rec}$. However, Fig. \ref{Fig12} indicates that whatever that effect on calibration accuracy is, it does not make redundancy algorithm unstable.

\begin{table*}[!ht]
\caption{Station configuration and the condition of redundancy calibration} 
\centering 
\renewcommand{\footnoterule}{} 
\begin{tabular}{l llllll} 
\hline\hline 
Station Conf.& no. of distinct redundant baseline& size($\mathbf{E}_{ph}$)& size($\mathbf{E}_{ampl}$)& $\kappa(\mathbf{E}_{ph})$& $\kappa(\mathbf{E}_{ampl})$& \textbf{$I$} \\[0.5ex]
\hline 
RS208 & 84 &$1127\times132$& $1125\times132$ & 52.72   & 9.74    & $99.65\%$\\[1ex] 
CS302 & 113&$1123\times161$& $1121\times161$ & 6.74e16 & 9.99    & $99.29\%$\\[1ex]
CS004 & 72 &$547\times120$ & $545\times120$  & 8.42e16 & 1.73e15 & $48.22\%$\\[1ex] 
\hline
\end{tabular} 
\label{tab:conf_cond}
\end{table*}

\section{Discussion}
\label{Disc.}

In this paper, we studied the applicability of redundancy calibration to phased array stations for the first time. Its performance was demonstrated using data of a new telescope, LOFAR. It required new considerations which were not part of the original design of conventional arrays e.g. WSRT and VLA. We took them into account by refining the data model. Reformulating the redundancy calibration formalism using the new data model, helped us to understand its potential, limitations and the effects of non-Gaussian baseline dependent noise on the calibration accuracy. 

Obtaining identical beams for station elements and minimizing the correlated noise, have to be considered as figures of merit for the configuration and spacings between the elements within a station in SKA (Square Kilometre Array) pathfinders and the SKA itself. This is definitely necessary, regardless of the calibration method to be used. For the redundancy method, it is fundamental, while having identical element beams saves computational capacity and time for model based methods.  

In the plots of residuals and Table. \ref{tab:cal_source}, we demonstrated how non-identicalness of the beams leads to systematic errors which can not be eliminated by any statistical methods or longer integration time. Since, identical element beams are fundamental especially for the redundancy calibration, similar EM-simulations to what we presented here, are highly recommended for any array which is to be calibrated using redundancy. Moreover, redundancy turns out to be an extremely suitable diagnostic tool for recognizing and monitoring failing station elements. This is of concern in arrays with large number of elements such LOFAR and SKA. It is currently being used for this purpose in the LOFAR system.

Wide field science with new radio telescopes such as SKA (\citet{2009IEEEP..97.1531V}), demands immense signal processing and computational capacity. This is mainly due to the large number of elements, variable beams and ionosphere over a large field of view. The redundancy calibration results encourage us to recommend to seriously consider redundancy in the SKA configuration, both at station and at whole array level. In large interferometers such as SKA, where the baselines are non-coplaner, one has to think of redundancy in UVW-space, however, not only in UV-space. Redundancy can be applied for gain calibration of the whole array in specific regimes where the stations observe the sky, through the same ionospheric patch. It also saves computational capacity and gives more accurate estimate of the telescope gains as compared to the model based gain calibration. This is of course a major step forward to achieve radio images of higher dynamic range. 

\section{Conclusion}
\label{Conclu.}

We studied for the first time the applicability and limitations of redundancy calibration in phased array stations using both real and simulated data of a radio telescope using the new aperture array technology, LOFAR. The results clearly show that the additional contraints from redundant baselines do improve the quality of the calibration and in addition provide a powerful tool for system diagnostics at different levels of the telescope phased array hierarchy (both intra station and inter station). The merit of redundancy in the station and full array layout and the additional advantage for diagnostic are well demonstrated by this study. We therefore strongly recommend to design for redundancy in both the station layout and the array configuration of future aperture arrays, in particular the SKA, where the required dynamic range will be an order of magnitude beyond any existing array. Because redundancy gives a better handle on characterizing the state of the system, it provides in addition a model independent diagnostic tool for subsystems such as a station.

\begin{acknowledgements}{The authors would like to heartfully thank Mariana Ivashina and Rob Maaskant for their generous and helpful discussions about antenna and system issues, Rob Maaskant and Michel Arts for
their help with CAESAR simulations, Sarod Yatawatta, Jan Noordam, Wim Brouw, Jaap Bregman and Ger de Bruyn for their very useful hints and discussions. They also thank the anonymous referee whose useful 
comments have improved the clarity of this paper considerably.}
\end{acknowledgements}

\bibliographystyle{abbrvnat}
\bibliography{RedunAandA}

\pagebreak

\appendix
 
\section{More plots of residuals for corrected redundant visibilities}
\label{appendix:a}

\begin{figure*}[!ht]
\centering
\includegraphics [width=15.5cm] {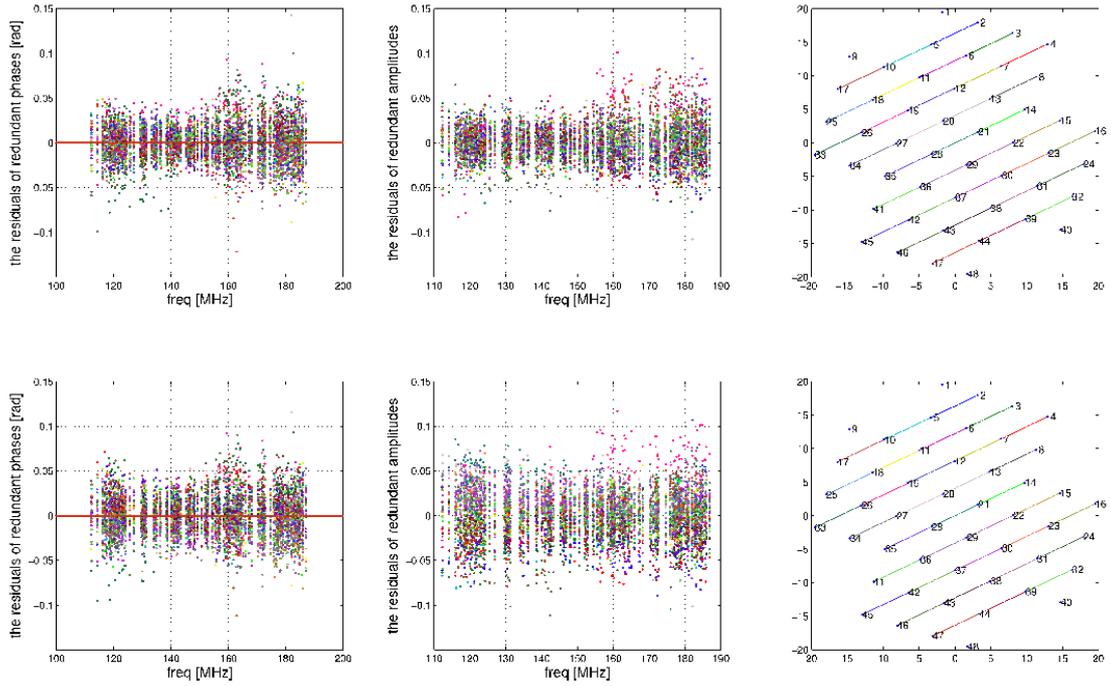}
\caption{Plots of the residuals for corrected redundant visibilities in terms of phases and amplitudes on a given set of redundant baselines. The first row shows the results after redundancy calibration. The second row shows the result afters model based calibration. The data is taken from an observation done on 24th November 2009 at 21:29:04 UTC, when CasA is at high elevation. The station configuration of RS208 is shown in the right most panel. The corresponding redundant baselines to each redundant visibility are also, with the same color code depicted on the station layout.}
\label{Fig_a1}
\end{figure*}

\begin{figure*}[!ht]
\centering
\includegraphics [width=15.5cm] {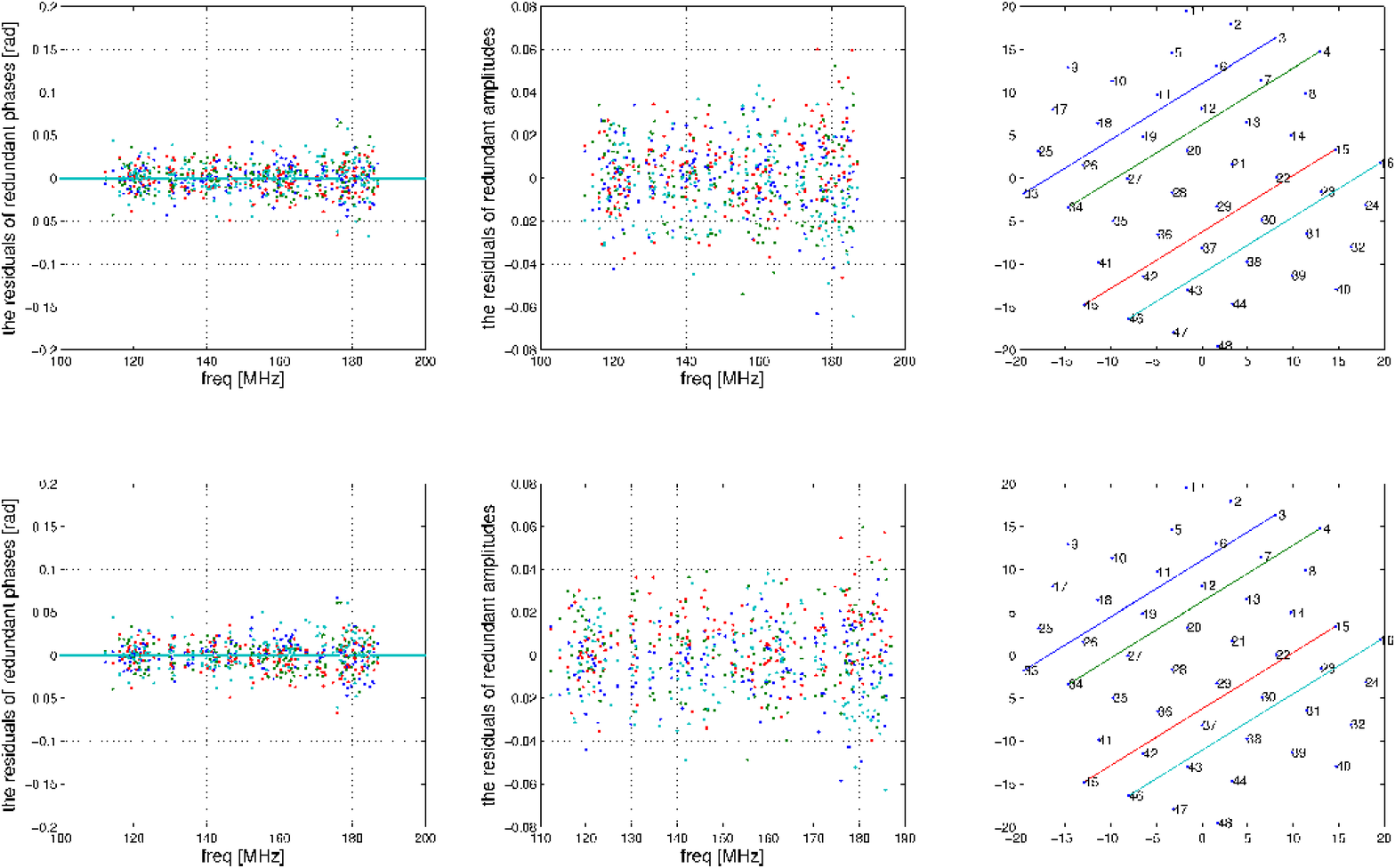}
\caption{Plots of the residuals for corrected redundant visibilities in terms of phases and amplitudes on a given set of redundant baselines. The first row shows the results after redundancy calibration. The second row shows the result afters model based calibration. The data is taken from the observation done on 24th November 2009 at 21:29:04 UTC, when CasA is at high elevation. The station configuration of RS208 is shown in the right most panel. The corresponding redundant baselines to each redundant visibility are also, with the same color code depicted on the station layout.}
\label{Fig_a2}
\end{figure*}

Some more plots of the residuals for the corrected redundant visibilities of our observational campaign on 24th November 2009 are presented here. Fig. \ref{Fig_a1} and Fig. \ref{Fig_a2} are similar results of 
the same data set (captured at 21:29:04 UTC, when CasA is at high elevation) presented in Fig.\ref{Fig9} but for different baselines, shown on the right most panel.

\begin{figure*}[!ht]
\centering
\includegraphics [width=15.5cm] {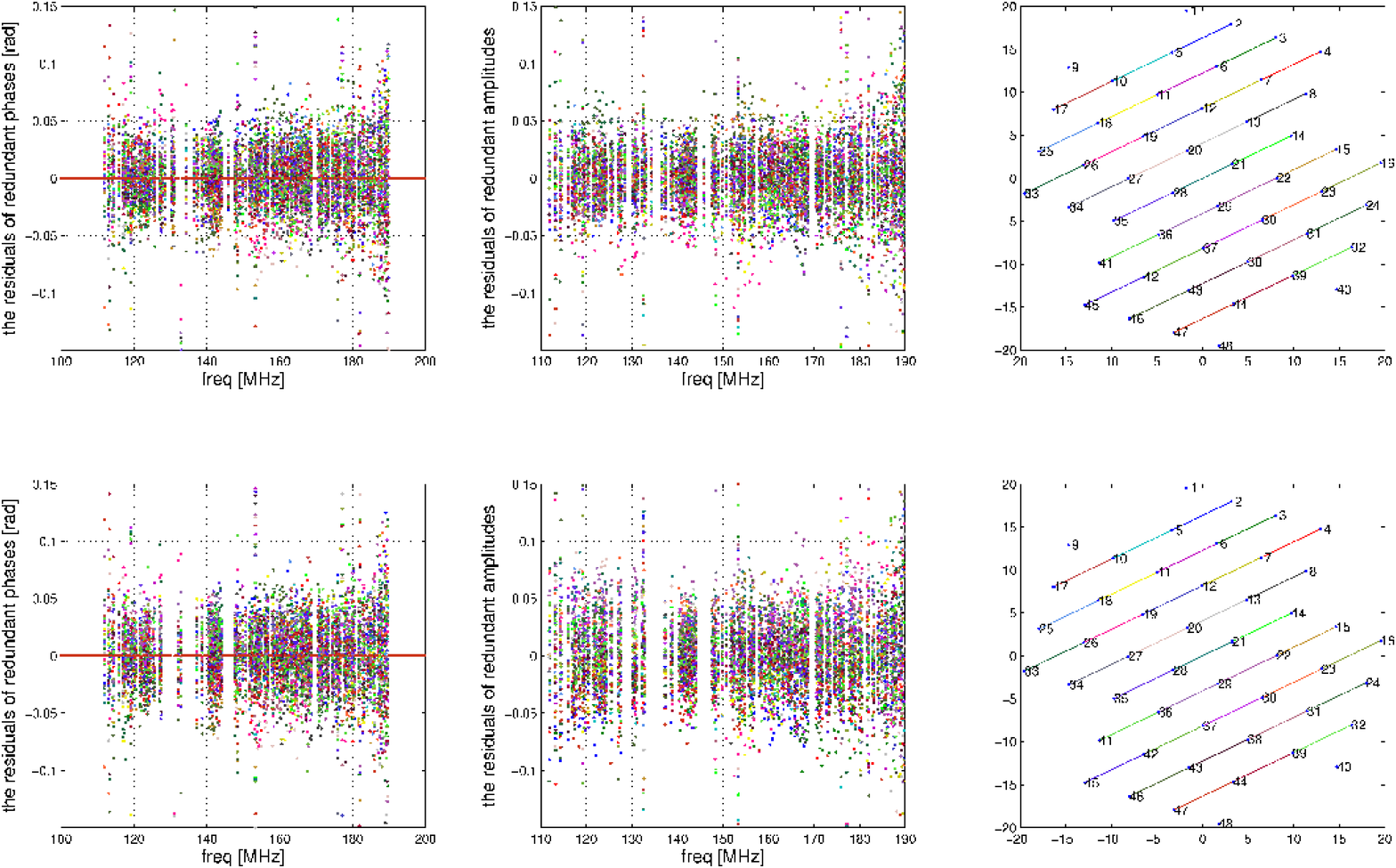}
\caption{The plots of the residuals for corrected redundant visibilities in terms of phases and amplitudes on a given set of redundant baselines. The first row shows the results after redundancy calibration. The second row shows the result afters model based calibration. The data is taken from the observation done on 24th November 2009 at 15:25:43 UTC, when CasA is  at low elevation. The station configuration of RS208 is shown in the right most panel. The corresponding redundant baselines to each redundant visibility are also, with the same color code depicted on the station layout.}
\label{Fig_a3}
\end{figure*}

\begin{figure*}[!ht]
\centering
\includegraphics [width=15.5cm] {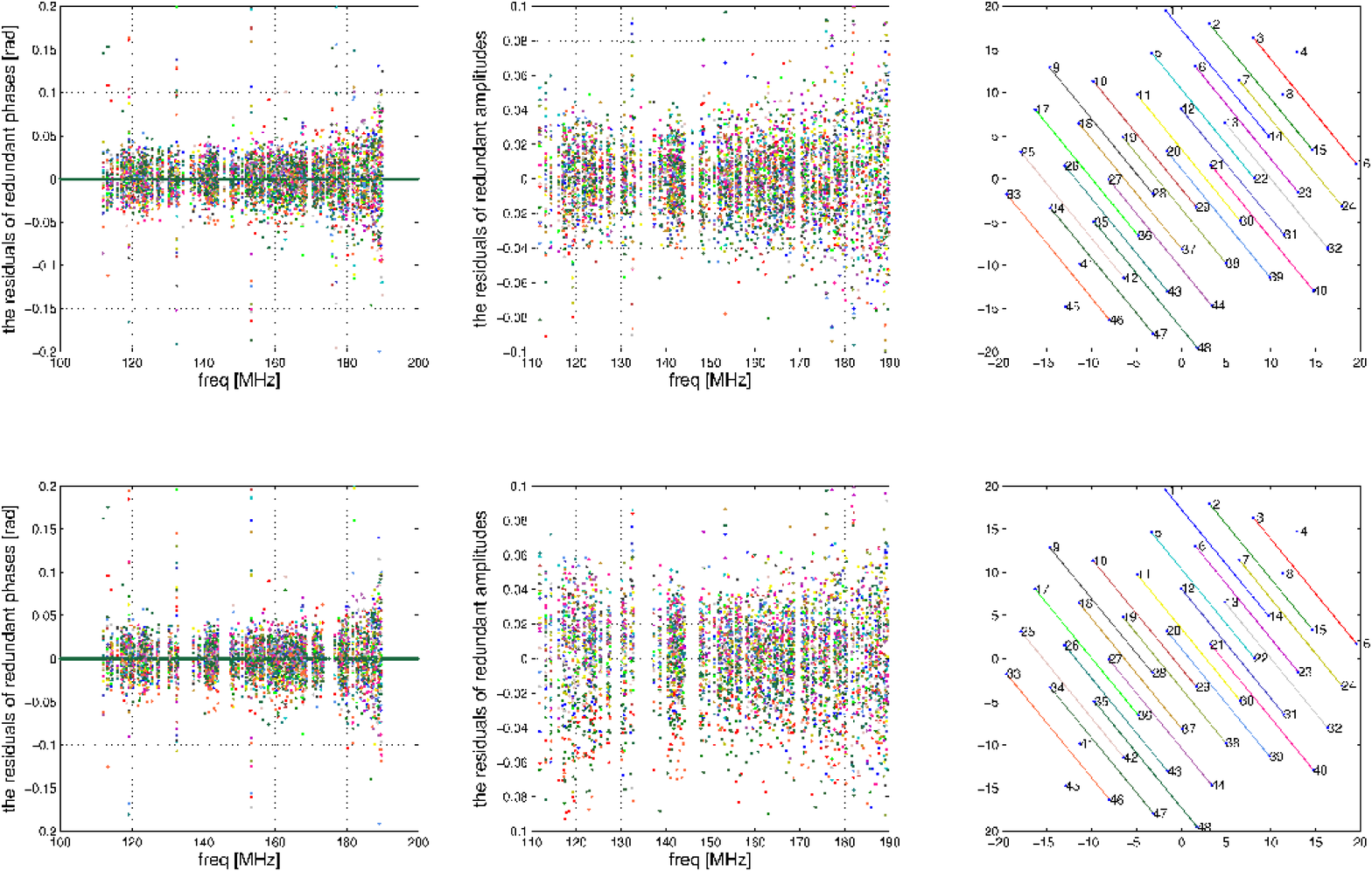}
\caption{The plots of the residuals for corrected redundant visibilities in terms of phases and amplitudes on a given set of redundant baselines. The first row shows the results after redundancy calibration. The second row shows the result afters model based calibration. The data is taken from the observation done on 24th November 2009 at 15:25:43 UTC, when CasA is at low elevation. The station configuration of RS208 is shown in the right most panel. The corresponding redundant baselines to each redundant visibility are also, with the same color code depicted on the station layout.}
\label{Fig_a4}
\end{figure*}

\begin{figure*}[!ht]
\centering
\includegraphics [width=15.5cm] {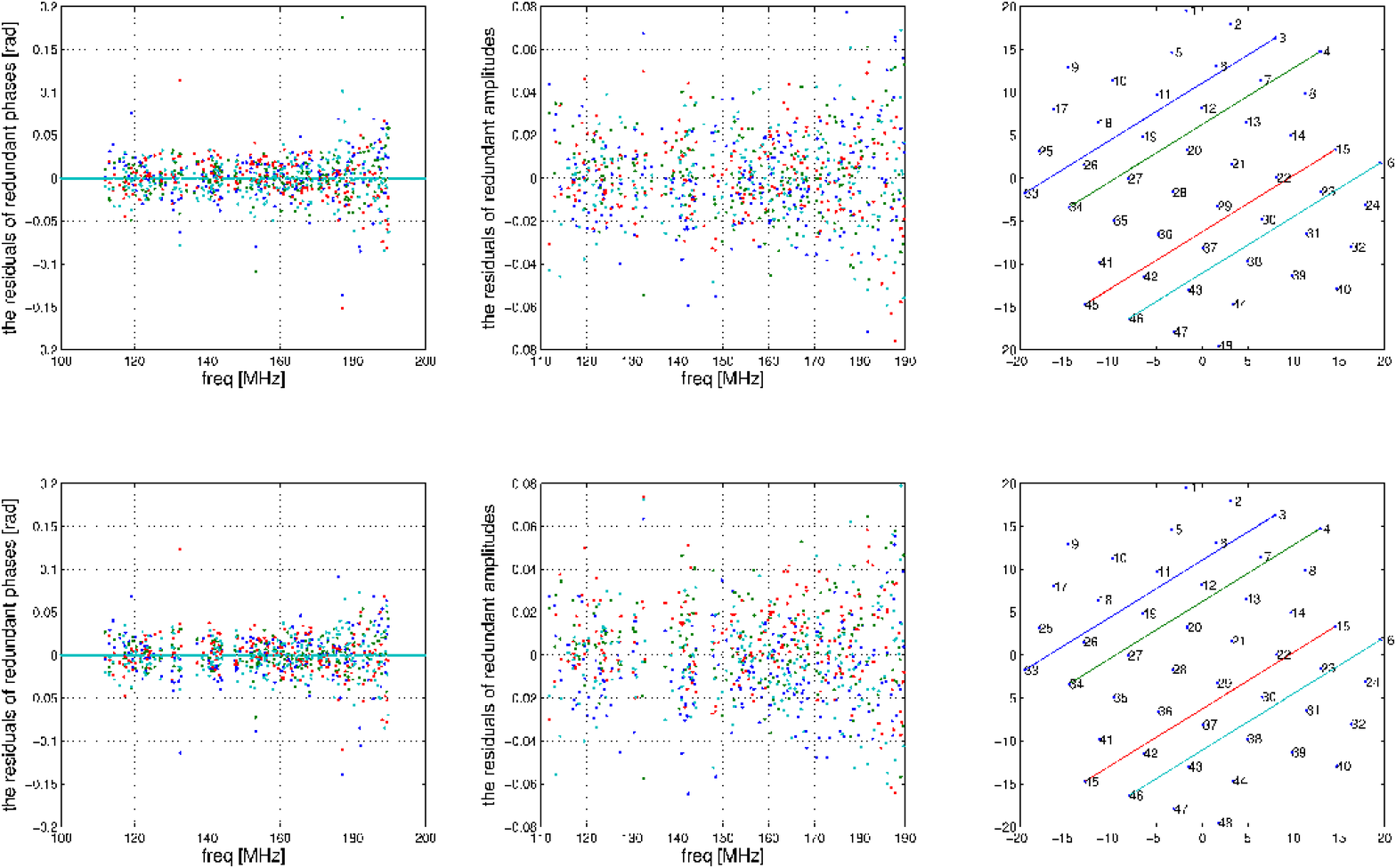}
\caption{The plots of the residuals for corrected redundant visibilities in terms of phases and amplitudes on a given set of redundant baselines. The first row shows the results after redundancy calibration. The second row shows the result afters model based calibration. The data is taken from the observation done on 24th November 2009 at 15:25:43 UTC, when CasA is at low elevation. The station configuration of RS208 is shown in the right most panel. The corresponding redundant baselines to each redundant visibility are also, with the same color code depicted on the station layout.}
\label{Fig_a5}
\end{figure*}

Fig. \ref{Fig_a3}- Fig. \ref{Fig_a5} show similar results using the data captured at 15:25:43 UTC, when CasA is at low elevation. They reveal slightly larger residuals in the order
of $5-10\%$ as compared with the results in Fig. \ref{Fig9}, Fig. \ref{Fig_a1} and Fig. \ref{Fig_a2}. This is due to a different mutual coupling environment which leads to less identical beams either in the main
lobe or the sidelobes. As it was discussed in Sect. \ref{CalQ.}, we still see smaller residuals after redundancy calibration as compared with the model based method.

\end{document}